\documentclass[%
 sor,
 jor,
 amsmath,amssymb,
 reprint,%
]{revtex4}

\usepackage {setspace}
\usepackage{graphicx}% Include figure files
\usepackage[english]{babel}
\usepackage{tabularx}
\usepackage{dcolumn}% Align table columns on decimal point
\usepackage{bm}% bold math
\usepackage{tikz}
\usepackage{graphicx}
\usepackage{amssymb}
\usepackage{amsmath}
\usepackage{bm}
\usepackage{enumitem}
\usepackage{etoolbox}
\begin{document}

\preprint{AIP/123-QED}

\title{Strain localization and yielding dynamics in disordered collagen networks}% Force line breaks with \\
%\thanks{Footnote to title of article.}

\author{Swarnadeep Bakshi$^a$}%
 %\affiliation{}%Lines break automatically or can be forced with \\
\author{Vaisakh VM$^{a,b}$}
%\affiliation{} %This line break forced with \textbackslash\textbackslash\\

\author{Ritwick Sarkar$^a$}
 %\homepage{http://www.Second.institution.edu/~Charlie.Author.}

\author{Sayantan Majumdar$^a$}
% \homepage{http://www.Second.institution.edu/~Charlie.Author.}
\thanks{smajumdar@rri.res.in}
\affiliation{$^a$Soft Condensed Matter Group, Raman Research Institute, Bengaluru 560080, India\\ $^b$Department of Physics, HKUST, Clear Water Bay, Hong Kong}

\date{\today}% It is always \today, today,

\begin{abstract}
Collagen is the most abundant extracellular-matrix protein in mammals and the main structural and load-bearing element of connective tissues. Collagen networks show remarkable strain-stiffening which tune the mechanical functions of tissues and regulate cell behaviours. Linear and non-linear mechanics of in-vitro disordered collagen networks have been widely studied using rheology for a range of self-assembly conditions in recent years. However, the correlation between the onset of macroscopic network failure and local deformations is not well understood in these systems. Here, using shear rheology and in-situ high-resolution boundary imaging, we study the yielding dynamics of in-vitro reconstituted networks of uncrosslinked type-I collagen. We find that in the non-linear regime, the differential shear modulus ($K$) of the network initially increases with applied strain and then begins to drop as the network starts to yield beyond a critical strain (yield strain). Measurement of the local velocity profile using colloidal tracer particles reveals that beyond the peak of $K$, strong strain-localization and slippage between the network  and the rheometer plate sets in that eventually leads to a detachment. We generalize this observation for a range of collagen concentrations, applied strain ramp rates, as well as, different network architectures obtained by varying the polymerization temperature. Furthermore, using a continuum affine network model, we map out a state diagram showing the dependence of yield-stain and -stress on the microscopic network parameters. Our findings can have broad implications in tissue engineering and designing highly resilient biological scaffolds.  
\end{abstract}

%\begin{keyword}
    %collagen, fracture, network, rheology
%\end{keyword}
 %Use showkeys class option if keyword
\maketitle

%\begin{quotation}
%The ``lead paragraph'' is encapsulated with the \LaTeX\ 
%\verb+quotation+ environment and is formatted as a single paragraph before the first section heading. 
%(The \verb+quotation+ environment reverts to its usual meaning after the first sectioning command.) 
%Note that numbered references are allowed in the lead paragraph.
%The lead paragraph will only be found in an article being prepared for the journal \textit{Chaos}.
%\end{quotation}

\section{Introduction}
Biopolymer networks, the major structural component of intra- and extracellular environment in animal body, show strikingly different mechanical properties compared to synthetic polymer gels \cite{Storm2005May, Janmey2007Jan, Chaudhuri2007Jan, LicupStressControlledMechanics, Stein2011Mar, Fratzl1998Jan, Gardel2004May, Broedersz2014Jul}. Properties like non-linear strain stiffening, negative normal stress in biopolymers are related to the `semi-flexibility' of the filaments as indicated by their relatively high bending rigidity ($\kappa$). In polymeric systems, the filament rigidity is generally expressed in terms of thermal persistence length ($L_p = \frac{\kappa}{k_B\,T}$) that indicates how tangent-tangent angular correlation decays along the filament due to thermal fluctuations. For biopolymers, $L_p$ is significantly larger than that for the synthetic polymers, yet, much smaller compared to a rigid rod. This indicates that despite of their rigidity, biopolymers show significant thermal bending fluctuations \cite{Onck2005Oct, Brangwynne2007Jul, Broedersz2014Jul}.
\newline
\newline
Type-I collagen is the most abundant protein in the extracellular matrix (ECM) of mammalian cells. Besides providing a scaffold for connective tissues, the mechanical and structural properties of ECM governs crucial cellular functions like cell proliferation, adhesion, migration, wound-healing, signaling etc \cite{BruceAlberts2018Aug, Rozario2010May, Mouw2014Dec, Wen2012Jan}. Abnormalities in ECM stiffness gives rise to various pathological conditions \cite{Conklin2011Mar, Walker2018Oct}.
\newline
\newline
To better understand the effect of linear and non-linear mechanics on various cellular and tissue functionalities, in-vitro reconstituted  assemblies of disordered isotropic collagen networks have become very popular in recent years, particularly, in the context of bio-physics, mechano-biology and tissue engineering \cite{Vashist2013Dec, Malda2013Sep, Sapudom2015Jun}. Such studies consider both simple shear \cite{Nam2016May, YANG2009Rheo} and extensional deformations \cite{Vader2009}. Non-linear mechanics of collagen is complex and highly architecture dependent \cite{JANSEN20182665}. Collagen networks with similar linear moduli can have widely different non-linear strain stiffening response \cite{Motte2013Jan}. Entropic elasticity of individual filaments, non-affine deformations, network heterogeneity, stress induced changes in network architecture contribute to the non-linear mechanics in these systems. Interestingly, collagen networks can show stability and finite elasticity for the average local network connectivity $\left\langle z \right\rangle$  varying approximately between 3 and 4 which is well below the isostaticity ($\left\langle z \right\rangle$ = 6 in three dimension) as predicted by Maxwell criterion \cite{Maxwell1864Apr}. Theoretical models demonstrate that bending rigidity of filaments stabilizes such sub-isostatic networks and strain stiffening originates from a bending to a stretching dominated response of the individual filaments/bundles \cite{LicupStressControlledMechanics, Mulla2019May, Sharma2016Jun}. Continuum unit-cell models (e.g. 3-Chain and 8-Chain Models) of semi-flexible filaments, provide compact analytical expressions effectively describing the non-linear strain stiffening in different biopolymer networks \cite{Meng, Palmer2008May}. 

Rheology and in-situ microscopy techniques like, confocal fluorescence (CFM), confocal reflectance (CRM) and boundary stress microscopy (BSM) provide important insight into the correlation of visco-elasticity, network deformation, failure and stress heterogeneity with the local network structure in these systems \cite{Tran-Ba2017Oct, Burla2020Apr, Arevalo2015Mar}. Stress relaxation in these systems strongly influenced by the magnitude of applied strain \cite{Nam2016May}. Similar to other biopolymers, non-linear mechanics of collagen networks also demonstrates striking hysteric effects \cite{Majumdar2018Mar, Schmoller2010Dec, Munster2013Jul}.  
\newline
\newline
Although, non-linear strain stiffening in disordered isotropic collagen networks has been extensively studied, much less attention has been paid to the dynamics of yielding and network failure. It is found that network architecture affects the magnitude of yield strain in collagen. Networks polymerized at higher temperature having finer fibrils with smaller pore sizes show a much larger yield strain compared to that corresponding to more bundled networks obtained at lower polymerization temperatures \cite{Motte2013Jan, JANSEN20182665}. Non-linear mechanics and yielding in these systems also show interesting system-size dependence over the length scales much larger than the network mesh size \cite{Arevalo2010Oct}. A very recent study  attempts to correlate the fracture strain with the local connectivity and plasticity of the collagen network using rheology and in-situ CFM. They directly probe the network failure over the length scales of single filaments to a few mesh size \cite{Burla2020Apr}. 
\newline
\newline
Despite of these studies, a direct correlation between the onset of bulk network failure and microscopic deformations is not fully understood. Moreover, in all the studies probing change in network structure using CFM and CRM in conjugation with rheological measurements, the imaging is done in the flow-vorticity plane compatible with standard microscopy set-up. Particularly, high resolution imaging in the flow-gradient plane is difficult due to the presence of air-sample interface in a plate-plate or cone-plate geometries used for the rheology measurements. Interestingly, the possibility of strain localization in the flow-gradient plane  has been speculated in the context of system-size dependent stiffening and yielding in collagen networks \cite{Arevalo2010Oct}. However, to our knowledge, in-situ strain distribution over mesoscopic to macroscopic length scales in flow-gradient plane has not been explored in these systems.
\newline 
\newline
Here, we study non-linear strain stiffening and yielding behaviour in uncrosslinked type-I collagen networks using rheology and in-situ  imaging over a range of concentrations and polymerization temperatures. Remarkably, for the first time, we probe the in-situ sample deformation and flow in the flow-gradient plane in collagen networks seeded with colloidal tracer particles. We find that the network softening or yielding starts well below the breaking/rupture strain. We also observe that high strain accumulates in the sample near the shearing boundaries when the applied strain crosses the yield strain. Such localized strain initiates a slippage between the network and the rheometer plate eventually leading to a detachment. Furthermore, fitting the stress-strain curves with a unit cell based network model for affine deformations, we map out a state diagram  connecting the yielding behavior with the reduced persistence length and mesh-size of the network.

\section{Results and Discussions}
We reconstitute networks of Type-I collagen starting from collagen monomers. Collagen monomers are formed by triple-helical polypeptides of length $\approx$ 300 nm and width $\approx$ 1.5 nm \cite{Gautieri2011Feb, Utiyama1973}. Under suitable buffer condition (Materials and Methods) the polymerization process takes place giving rise to space-filling collagen networks. Polymerization time and network architecture strongly depends on the parameters like temperature \cite{RAUB20072212, Wolf2013Jun, Hwang2011Mar}, pH \cite{Raub2008Mar, Roeder2002Apr, Harris2007Jul}, ionic concentration \cite{Harris2007Jul, Lang2015Feb}. The typical network architecture for two different polymerization temperatures are shown in Fig. 1(a) and (b). At 4$^{o}$C, the network is very heterogeneous  with thick parallel bundles, whereas, more homogeneous network with finer fibrils are observed at 25$^{o}$C. The distributions of fibril/bundle diameter for different temperatures are obtained from the freeze fracture SEM data (sample size: N $\sim$ 400) is shown in Fig. 1(c). We find that over the temperature (T) range of 4$^{o}$C - 35$^{o}$C, both the mean and standard deviation filament diameter decreases with increasing temperature, as also reported in earlier studies \cite{RAUB20072212, YANG20092051, JANSEN20182665}.
\begin{figure*}
    \begin{center}
    %trim from left edge
    \includegraphics[height = 9 cm]{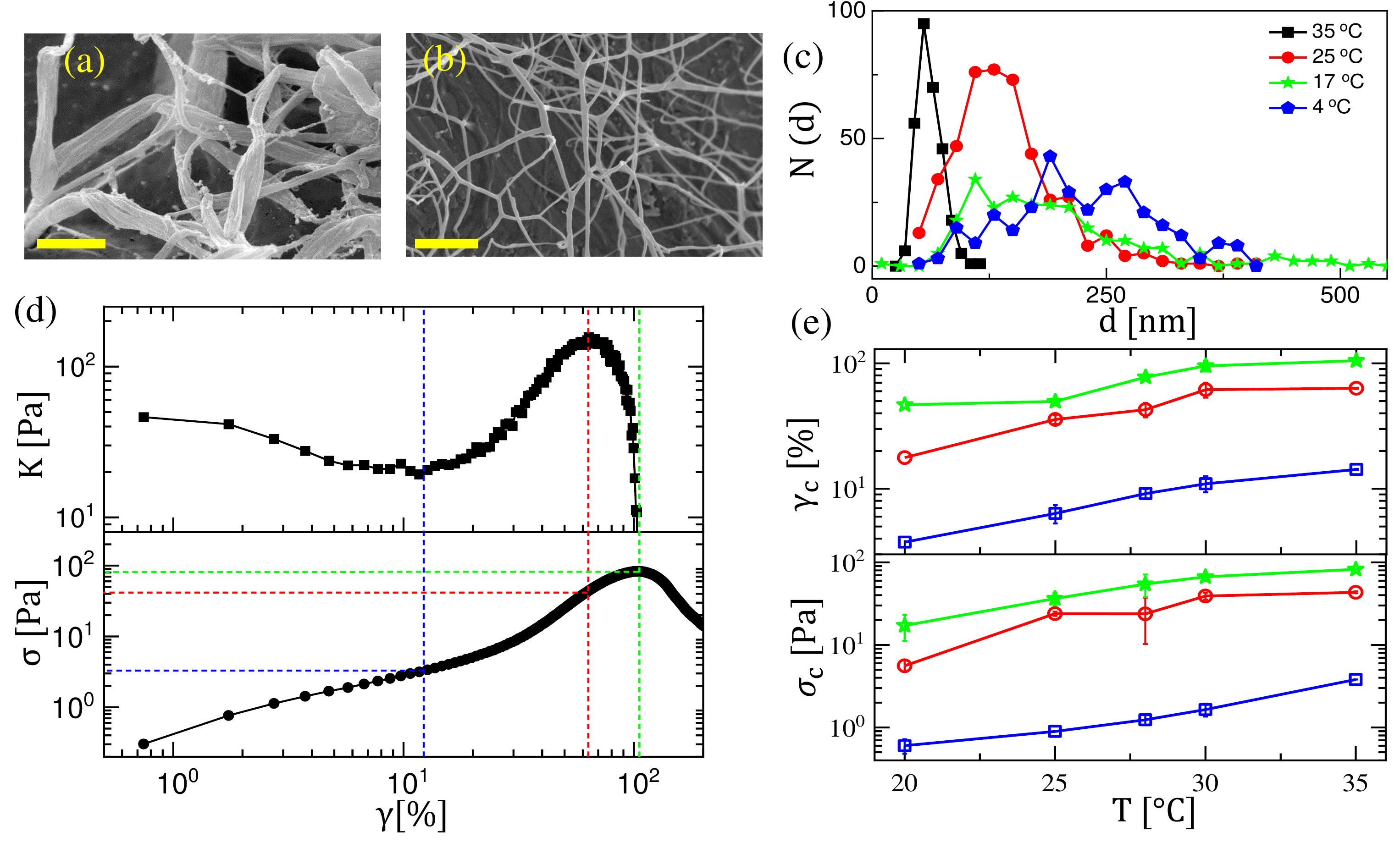}
    %trim from right edge
    %\includegraphics[]{F1.PDF}
    \caption{Freeze-fracture SEM image of collagen polymerized at (a) 4$^{o}$C (Scale = 1 $\mu$m) and (b) 25$^{o}$C (Scale = 2 $\mu$m). (c) Bundle fibril diameter distribution. For Fig. 1(a),(b) and (c), the collagen concentration $\phi$ = 1 mg/ml. (d) Variation of differential shear modulus (top panel) and stress  (bottom panel) with applied strain for polymerization temperature T = 35$^{o}$C. Critical stress and strain points corresponding to the onset, yield and breaking are marked with the dashed lines. (e) Variation of critical strain and stress as a function of polymerization temperature (squares: onset stress/strain, circles: yield stress/strain and stars: breaking stress/strain). Error bars represent standard deviations for two independent measurements. For Fig. 1(d) and (e) the collagen concentration $\phi$ = 2 mg/ml. }
    \label{F1}
    \end{center}
\end{figure*} 

Rheology measurements are carried out on a  MCR-702 stress-controlled rheometer (Anton Paar, Graz, Austria) using cone-plate geometry having rough sand-blasted surfaces (Materials and Methods). Such rough surfaces minimize wall-slippage of the sample under shear. The polymerization process of collagen is monitored by applying a small oscillatory shear strain (amplitude $\gamma_0$ = 2\%, frequency $f$ = 0.5 Hz) and measuring the linear visco-elastic moduli $G'$ and $G''$ as a function of time. After an initial increase, $G'$ and $G''$ reaches a plateau value (Fig. S1), indicating that the network is polymerized. We carry out the rheology measurements over a concentration ($\phi$) range of 1 mg/ml - 3 mg/ml. In all cases, we find that over a wide range of frequency $G'(f)$ is much larger than $G''(f)$ (Fig. S2), indicating that the networks behave like a visco-elastic solid. To probe the response of the network as a function of strain, we apply a constant strain ramp rate $\dot{\gamma}$ = 1\%/s on the network and measure the stress response. We show the variation of shear stress ($\sigma$) and differential shear modulus $K = \frac{d\sigma}{d\gamma}$ as a function of $\gamma$ (Fig. 1(d)) for $\phi$ = 2 mg/ml and T = 35$^{o}$C. We find that after a mild strain-weakening regime, $K$ increases rapidly beyond an onset strain $\gamma_o$ indicating the non-linear strain stiffening of the network, when the shear modulus of the network increases with increasing strain. At larger strain values, $K$ reaches a maximum before starting to drop beyond $\gamma_y$, the yield-strain of the network, indicating a network weakening under large strain. Interestingly, the peak stress is reached at a higher strain value compared to $\gamma_y$. We define the strain value at which the stress reaches the peak as the breaking/rupture strain $\gamma_b$ for the network.  Beyond $\gamma_b$ the stress weakening starts. Below yielding the network deformation is reversible, however beyond yielding/rupture irreversibility sets in (Fig. S3). Mathematically, $\frac{d^2\sigma}{d\gamma^2}|_{\gamma_y}$ = 0 and $\frac{d\sigma}{d\gamma}|_{\gamma_b}$ = 0 defines the yielding and breaking points of the network, respectively. From Fig. 1(d) and (e) we note that $\gamma_y \approx 50\%$ where as, $\gamma_b \approx 100\%$. This points out that for disordered collagen networks yielding is a gradual process rather than an abrupt one. Similar strain response is also observed for cone-plate geometry with smooth surfaces, but the measured value of the shear modulus ($K$) remains considerably lower compared to that obtained from the rough geometry (Fig. S4). We show the variation of $\gamma_o$, $\gamma_y$ and $\gamma_b$ and the corresponding stress values ($\sigma_o$, $\sigma_y$, $\sigma_b$) as a function of polymerization temperature in Fig. 1(e) (top and bottom panels). We find that both the critical strain and stress values increase with increasing temperature. Similar trends are also observed for $\phi$ = 1 mg/ml and 3 mg/ml (Fig. S5). For temperatures below 20$^{o}$C, we observe condensation of small water droplets on the rheometer plates. This can change the moment of inertia of the shearing plates and introduce measurement artifacts. Thus, for rheology measurements we do not probe any temperature below 20$^{o}$C. To check the robustness of the yielding behaviour, we also study the sample response using a pre-stress protocol extensively used in the bio-polymer literature \cite{Broedersz2010Aug}. We observe a stress-stiffening behaviour when the tangent storage ($K'$) and loss ($K''$) moduli increase with the increasing pre-stress magnitude (Fig. S6). However, for larger pre-stress values, there is an abrupt drop in $K'$ and $K''$ indicating sample weakening/yielding. The stress stiffening and yielding correlates well with that obtained under steady-shear (Fig. S6). 

\begin{figure*}
    \begin{center}
    %trim from left edge
    \includegraphics[height = 5.5 cm]{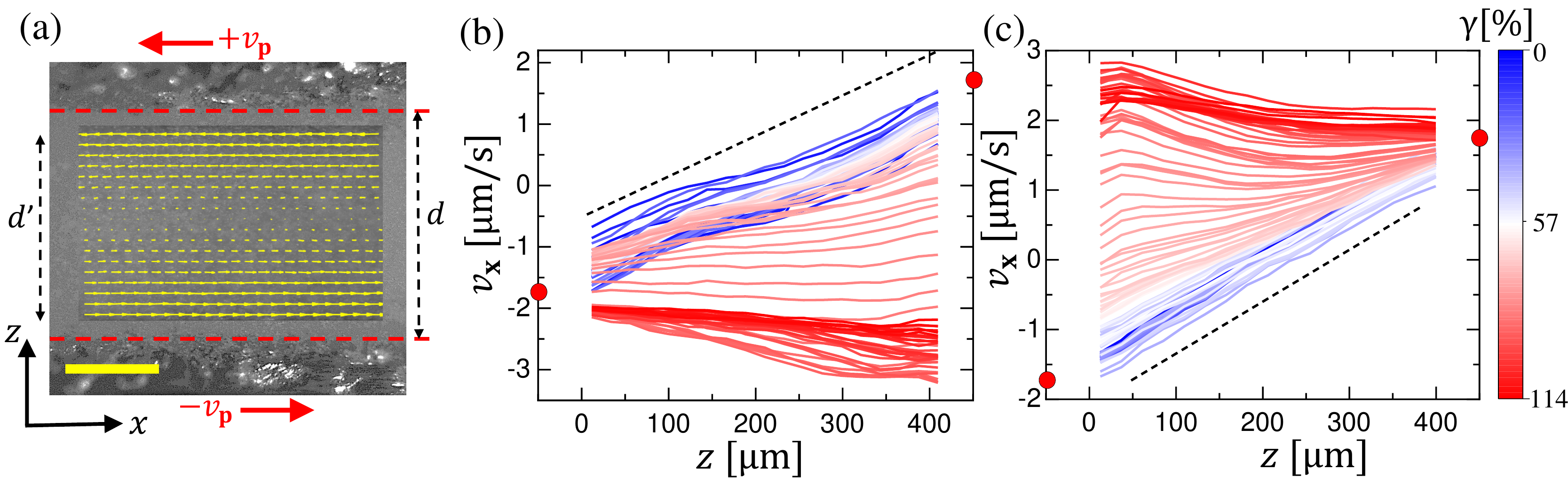}
    %trim from right edge
    %\includegraphics[]{F1.PDF}
    \caption{(a) Typical boundary image with superposed velocity profile for the collagen sample seeded with 2.8 $\mu$m polystyrene particles (1\% v/v). $d$ and $d'$ respectively represent the gap between the plates and the sample width used for PIV analysis. $\pm v_p$ denote the plate velocities and the scale bar represents 200 $\mu$m.	(b) and (c) show the velocity profiles  obtained from the PIV analysis for T = 30$^{o}$C and 25$^{o}$C, respectively. Color gradient represent increasing applied strain ($\gamma$) with a ramp rate of 1\%/s. The red dots indicates the plate velocities. The dashed lines represent the velocity profile predicted assuming affine deformations. We see strong non-affine deformation for larger strain values close to yielding.
		Here, $\phi$ = 2 mg/ml.}
    \label{F2}
    \end{center}
\end{figure*}

To get a deeper insight into the yielding behaviour of collagen networks, we map out the spatio-temporal evolution of the velocity field in the sample  using particle imaging velocimetry (PIV) technique. As shown in the experimental set-up (Fig. S7) we illuminate the sample using a LED light source (Dolan-Jenner Industries) and image the diffused scattering from the sample boundaries in the flow-gradient plane using a digital camera (Lumenera) fitted with a 5X, long working distance objective (Mitutoyo). We do not get any appreciable scattering from the pure collagen and the sample appears almost transparent. We put 1\% (v/v) polystyrene tracer particles ($d$ = 2.8 $\mu$m) \cite{Dhar2019} in the collagen samples to enhance the scattered intensity required for the PIV measurement (Fig. 2(a)). We find that $\sim$ 1\% (v/v) is the minimum amount of tracer particles required to get a fairly uniform distribution of speckle pattern. Freeze fracture SEM images (Materials and Methods) show that the particles have some affinity to stick to the networks (Fig. S8), however, FTIR spectra (Materials and Methods) points out that no chemical bonds between the particles and collagen fibers are formed (Fig. S9). We also confirm that introducing such low concentration of tracer particles does not modify the rheological behaviour of collagen networks (Fig. S10). A typical boundary image displaying the speckle pattern with velocity vectors obtained from PIV analysis is shown in Fig. 2(a). The plates are moving with velocities $+v_p$ and $-v_p$ in a counter-rotation configuration giving an applied shear rate $\dot{\gamma} = \frac{2v_p}{d}$, where $d$ is the gap between the plates. Evolution of the spatially-averaged velocity profiles across the gap [$v_x(z)$ vs $z$] with increasing strain values are shown in Fig. 2(b) and 2(c) for polymerization temperatures of 30$^{o}$C and 25$^{o}$C, respectively. We see that for all strain values $\gamma < \gamma_y$, the profiles remain almost linear, indicating an affine deformation. The average shear rate inside the sample below $\gamma < \gamma_y$ is given by $\frac{2v_0}{d'}$, where $\pm v_0$ indicate the velocities inside the PIV window at $z = 0$ and $z = d'$ (Fig. 2(a)). However, for  $\gamma \geq \gamma_y$ the velocity profiles evolve rapidly, in particular, near one of the plate boundaries (Fig. 2(b) and 2(c)) and finally a detachment happens at $\gamma = \gamma_b$ (also see Movie 1). Such evolution of velocity profile  indicates strain localization and slippage near the boundary. Beyond the detachment, the shear strain in the network becomes negligible and entire network moves with a constant velocity with the plate remains attached to. Sometimes, the network can also detach from both the plates (Movie 2). Interestingly, for few rare occasions we observe that some part of the network can reattach with the moving plates after the initial detachment giving rise to more complex fracture patterns (data not shown). The boundary near which velocity profiles rapidly evolve and finally sample detachment takes place, randomly varies for different experimental runs and most likely depends on the attachment of the network with the rheometer plates. Such trend is observed for all sample concentrations, temperatures and applied shear rates we consider in the present study. Our observations indicate that the yielding onset of collagen networks under shear is governed by boundary dynamics. Interestingly, at lower strain values (strain softening region) where the velocity profiles are almost linear, we observe small up/down shifts of the entire profile as shown in Fig. 2(b) and 2(c). This indicates that some rearrangements are going on near the boundaries even during the strain softening response.

\begin{figure*}[h]
    \begin{center}
    %trim from left edge
    \includegraphics[height = 6.2 cm]{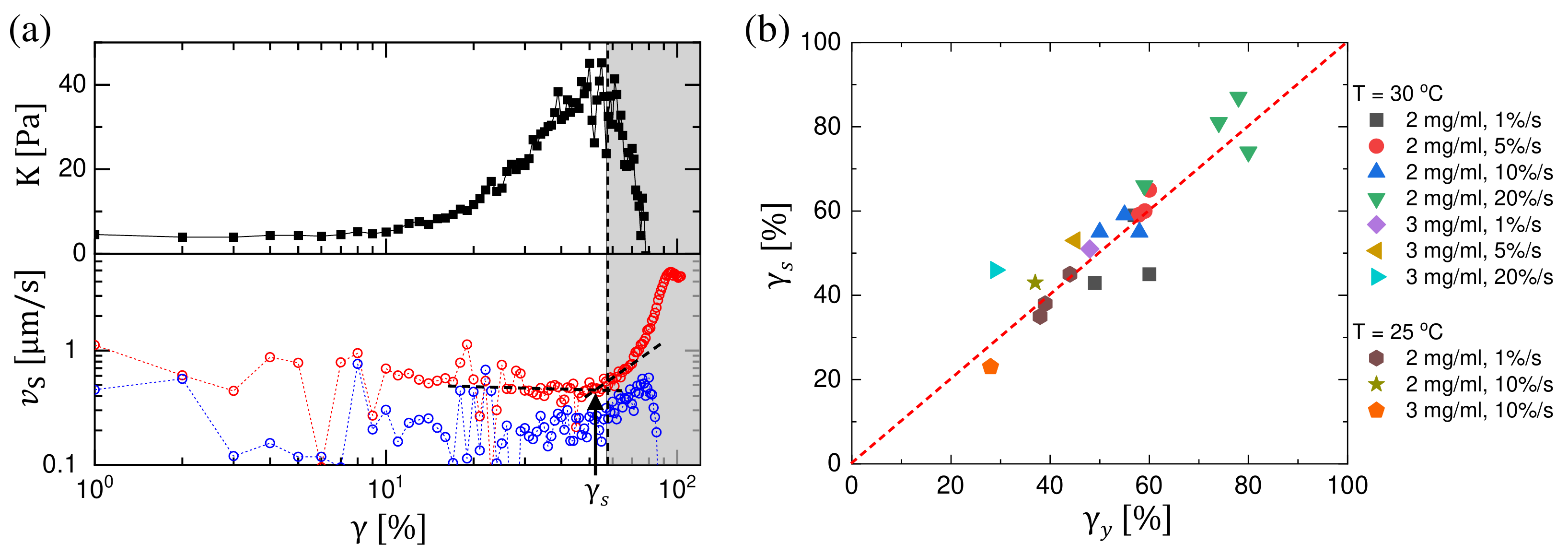}
    %trim from right edge
    %\includegraphics[]{F1.PDF}
    \caption{ (a) Variation of shear modulus $K$ (top panel) and slip velocity $v_s$ obtained from PIV analysis (bottom panel) for $\phi$ = 2 mg/ml and T = 30$^{o}$C as a function of applied strain. Slippage between the sample boundary and the rheometer plate show a significant increase beyond the yield strain ($\gamma_y$), as shown by the data in the shaded region. Otherwise, such slippage is negligible. The dashed vertical line represents $\gamma = \gamma_y$. The dashed lines in the bottom panel are guides to the eye representing the values of $v_s$ just before and after the slippage. The strain corresponding to the point of intersection of these lines gives the onset strain $\gamma_s$ beyond which significant slippage is observed. We mark $\gamma_s$ by an arrow in the bottom panel. (b) Variation of $\gamma_s$ as a function of $\gamma_y$ for a range of sample conditions and strain ramp rates, as marked in the legend. The dashed red-line corresponds to $\gamma_s = \gamma_y$.}
    \label{F3}
    \end{center}
\end{figure*}
\begin{figure*}
    \begin{center}
    %trim from left edge
    \includegraphics[height = 8 cm]{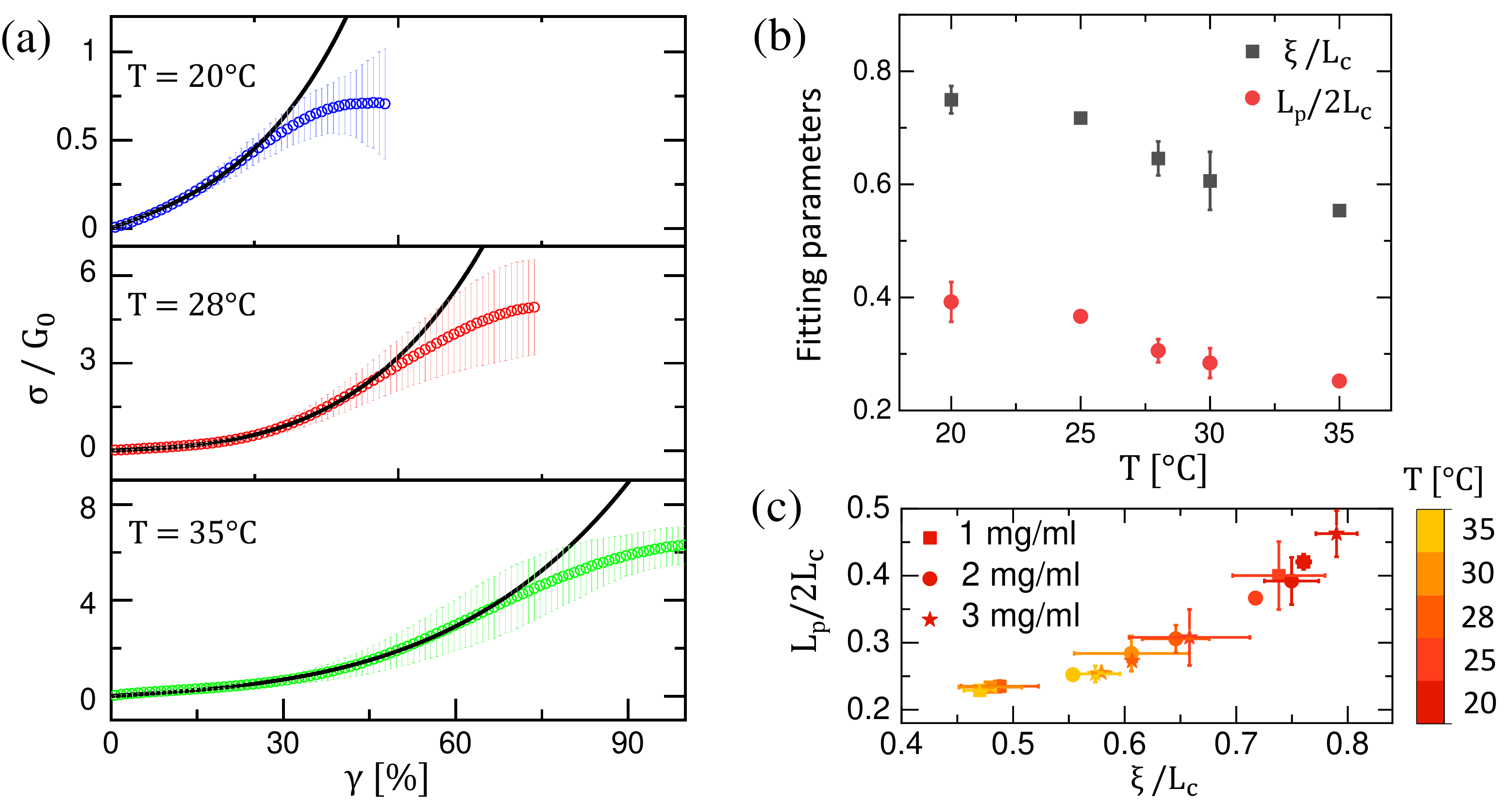}
    %trim from right edge
    %\includegraphics[]{F1.PDF}
    \caption{(a) Symbols indicate the variation of normalized stress as a function of applied strain for different polymerization temperatures as shown.  Solid lines represent the fits to the 8-chain model as described in the main text. We get very good agreement in all cases upto intermediate strain values, however, significant deviation from the fitted model is observed due to non-affine deformations close to yielding. (b) Variation of reduced persistence length ($L_p/2\,L_c$) and mesh size ($\xi/L_c$)  as a function of temperature obtained from the fitting for $\phi$ = 2 mg/ml. (c) $L_p/2\,L_c$ vs $\xi/L_c$ for different collagen concentrations and polymerization temperatures as indicated. The error bars in all the panels represent the standard deviations obtained from two independent experiments under the same sample condition.}
    \label{F4}
    \end{center}
\end{figure*}

To further quantify the boundary-failure dynamics of the network, we define the slip-velocity $v_s = |v_p - v_0|$ and plot it as a function of applied strain ($\gamma$) in Fig. 3(a) for T = 30$^{o}$C. We find from Fig. 3(a) that $v_s$ remains small as a function of $\gamma$ till $\gamma = \gamma_y$, for both the top and bottom plates. This indicates that, over a range of strain values, both in the linear and non-linear regime, the slippage between the sample and the plates is not significant. As $\gamma$ crosses $\gamma_y$ (indicated by the peak of $K$), $v_s$ starts to increase rapidly, particularly, for one of the plates and reaches a maximum near $\gamma = \gamma_b$. The maximum value of $v_s \sim 2 v_p$, however, in some cases the value of slip-velocity can be even larger due to strong elastic retraction of the network after the rupture. The applied strain value beyond which $v_s$ increases rapidly defines the slip-strain ($\gamma_s$), as shown in Fig. 3(a) (bottom panel). To establish the relation of boundary slippage with the yielding behaviour, we plot $\gamma_s$ as a function of $\gamma_y$ in Fig. 3(b) for varying sample conditions as well as, strain ramp rates. Remarkably, we observe a strong correlation between $\gamma_s$ and $\gamma_y$, further confirming that boundary slippage gives rise to yielding in collagen networks. 

Our observation points out the generality of boundary failure dynamics in governing the yielding behaviour of the networks over a wide range of parameters. Moreover, the appearance of significant boundary slippage only deep inside the strain-stiffening region indicates  that such detachment is triggered by internal stresses generated in the system due to strong non-linear deformation of the network. We indeed observe a significant negative normal stress in the system before yielding (Fig. S11). Currently, we do not fully understand why the network failure/rupture always takes place close to the sample boundaries. There can be a possible connection with network rarefaction under non-linear deformation as observed for highly cross-linked actin networks \cite{Schmoller2010Dec}. Such rarefaction results in lower average number of contacts between the filaments close to the boundaries giving rise to a local weakening of the sample. For some cases, we observe a drop in the scattered intensity from the sample near the shearing plates close to yielding (Fig. S12). Such intensity drop near the plates increases further beyond the breaking strain. This observation coupled with the fact that tracer particles have an affinity to stick to the collagen fibers (Fig. S8), also supports the local rarefaction picture mentioned earlier. However, more studies are required to confirm such deformation induced network heterogeneity in collagen systems.

\begin{figure*}
    \begin{center}
    %trim from left edge
    \includegraphics[height = 5.5 cm]{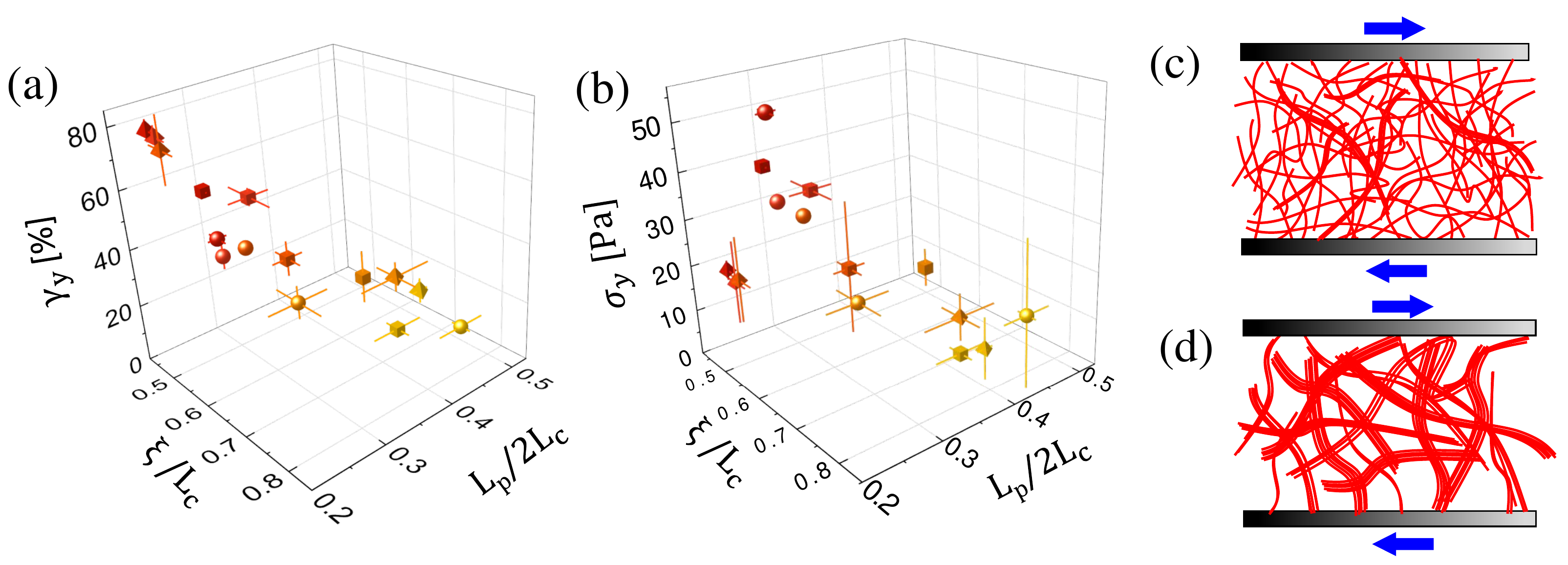}
    %trim from right edge
    %\includegraphics[]{F1.PDF}
    \caption{Variation of (a) yield strain ($\gamma_y$) and (b) yield stress ($\sigma_y$) with reduced persistence length and mesh size. Tetrahedron, cube and sphere shaped symbols correspond to 1 mg/ml, 2 mg/ml and 3 mg/ml collagen concentrations, respectively. Dark to light color represents the decreasing magnitude of $\gamma_y$ or $\sigma_y$. Error bars represent the standard deviations of two independent measurements. Here, $\phi$ = 2 mg/ml. Schematics showing the network structures obtained at a higher (panel (c)) and a lower (panel (d)) polymerization temperatures. Due to smaller mesh size the networks polymerized at a higher temperature have much larger number of contracts  with the plates compared to more bundled networks obtained at a lower temperature for same concentration of collagen monomers.}
    \label{F5}
    \end{center}
\end{figure*} 

Next, we turn to the temperature dependent microscopic network architecture that controls the yielding behaviour in these systems. Although, the mechanical properties of a single biopolymer filament is well described by Worm Like Chain model (WLC) \cite{Kratky1949Jan}, a network formed by many such filaments have random structures that make the derivation of analytical expression for free energy and stress-strain relation extremely complex \cite{Meng}. To simplify this problem, several lattice-based models have been proposed. These models essentially use the concept of a `mesh cell' or, an elemental volume, such that, the entire sample can be constructed just by 3-D translation of this elemental volume. The main assumption of this model is affine or uniform deformation in the system spanned by the deformation of the `mesh cell'. As mentioned earlier, we observe affine deformations over a wide range of strain values below yielding. Thus, such lattice based models should capture the mechanics for collagen networks for $\gamma < \gamma_y$. 
One such models for the `mesh cell' is \textit{8-chain model}, where the cell is body-centered cubic. Eight chains connect the eight corners of the cube to the center. The mesh size ($\xi$) of the network is given by the edge-length of the cubic cell.     
The stress ($\sigma$) vs strain ($\gamma$) relationship for this network model under simple shear is given by \cite{Meng}, 
\begin{equation}
\sigma = \frac{2}{3}nk_B T\gamma a^2\left[\frac{9}{c\pi[3-(3+\gamma^2)a^2]^2}-c\pi^2\right]
\end{equation}
where, the parameters $a=\frac{\xi}{L_c}$ and $c=\frac{L_p}{2L_c}$ are the reduced mesh-size and persistence length, respectively.  Here,  $n$ is the  number of entanglement points per unit volume and $L_c$ is the filament contour length. A dimensionless form of the above equation can be obtained if we divide both sides by the average plateau modulus ($G_0 = \sqrt{G'^2 + G''^2}$) obtained from the frequency sweep data (Fig. S2). We use the above equation to fit $\sigma / G_0$ vs $\gamma$ data obtained experimentally for different temperatures and collagen concentrations. In 8-chain model, the network is supposed to be comprised of individual filaments. However, in entangled biopolymers, bundle formation is inevitable \cite{Goh1998} (also see, Fig. 1(a) - 1(c)). Since, the bundles are aggregates of few individual filaments, in principle the 8-chain model can be extended for bundles assuming them as thicker filaments with larger $L_p$ values. However, an important difference between semi-flexible filaments and un-crosslinked bundles lies in their internal structure. When a bundled network deforms, the bundles not only deform as a whole but there are changes due to sliding of filaments inside individual bundles. These intra-bundle shearing dynamics makes the scenario very complex \cite{Heussinger2007Jul, Bathe2008Apr}. To have an essence of the microscopic architecture of the network, we use the 8-chain model where the lateral dimension of the polymer bundles is neglected and they are treated as chains (albeit with higher rigidity). Also, 8-chain model can not capture the strain softening behaviour at smaller strain values coming predominantly from the filament buckling \cite{Onck2005Oct}. For clarity, we show the fitted data only for $\phi$ = 2 mg/ml concentration at three different temperatures in Fig. 4(a). We see very good agreements for strain values $\gamma < \gamma_y$ in all cases. For the fitting initialization, we use $n$ values similar to those obtained in the  recent simulations for collagen networks \cite{Valero2018} (also see Table of Parameters in S.I.). The parameters $\frac{L_p}{2L_c}$ and $\frac{\xi}{L_c}$ obtained from the fitting are plotted in Fig. 4(b) as a function of polymerization temperature. We find that both of these parameters decrease with increasing temperature. This result also agrees with the variation of filament diameter obtained from the SEM images, where we observe that fibril thickness decreases with increasing temperature (Fig. 1(c)). The strong correlation between the parameters (Fig. 4(c)) implies bundle formation: for a given monomer concentration, presence of thick bundles will increase the average persistence length and mesh size of the network (see Fig. 5(c) and 5(d)). Importantly, from 8-chain model, we obtain the variation of reduced persistence length/mesh size less than a factor of 2 as shown in Fig. 4(b) for a temperature variation of 20$^0$C - 35$^0$C. However, the absolute values of $L_p$ and $\xi$ vary over a much wider range, since, for a bundle of diameter $d$ (assuming a cylindrical cross-section), $L_p = L'_p (\frac{d}{d_0})^4$, where $d_0$ and $L'_p$ are the mean diameter and persistence length of a single collagen filament. By measuring that average value of $d_0$ ($\approx$ 46 nm) from the SEM data that matches closely with the literature values \cite{Lodish2000} and using $L'_p$ = 10 $\mu$m \cite{Lovelady2014Apr}, we find that $L_p$ varies over a wide range (Fig. S13a) similar to that observed for bundled actin networks \cite{Gardel2004May}.  To be consistent with the 8-chain model, $L_c$ requires to also vary with temperature (Fig. S13b). However, further work is required to understand a physical meaning of $L_c$ in the 8-chain model for bundled networks. Furthermore, from the fitting parameters (see Table of Parameters in S.I.) we find that the ratio of the reduced persistence length and mesh size ($\frac{L_p}{2\xi} $) remains close to 1, as also reported for a wide range of biopolymers \cite{Meng2017Feb}. This implies that the network mesh size ($\xi$) should also be very large for a bundled network, much larger than the reported values in the literature for similar collagen concentrations obtained from confocal reflectance microscopy \cite{Arevalo2010Oct, Zhu2014Apr} and turbidity measurements \cite{JANSEN20182665}. Such discrepancy can arise from the fact that unlike dry SEM imaging, freeze-fracture SEM produces clear image of the networks only over a field of view much smaller than the persistence length ($L_p$) of collagen bundles. Thus, an actual bundle can be very non-uniform over the length scale of $L_p$. Since, $L_p \sim d^4$, presence of thinner regions in a bundle can act as local weak spots and drastically reduce the overall persistence length of the bundle.

We show the dependence of yield strain/stress ($\gamma_y$ or $\sigma_y$) on microscopic network parameters (reduced mesh-size and persistence length) by generalized three-dimensional state diagrams in Fig. 5(a) and (b). As discussed earlier, in our case, these parameters are highly correlated due to bundle formation. The variation of microscopic network architecture is obtained by varying the polymerization temperature (20$^0$C - 35$^0$C) as well as collagen concentration (1 mg/ml - 3 mg/ml). We see from Fig. 5(a) that with increasing value of reduced mesh size and persistence length $\gamma_y$ monotonically drops. This is remarkable, since, it indicates that despite of the complex network architecture the yield strain is essentially governed by two correlated microscopic network parameters related to the mesh size and persistence length. Networks with finer fibrils are more resilient. Similar trend is also observed for $\sigma_y$ (Fig. 5(b)). We note that, although, $\gamma_y$ shows similar dependence on the reduced mesh size and persistence length for all concentrations, $\sigma_y$ values for 1 mg/ml collagen networks are much lower. This observation is in line with the results mentioned in \cite{Burla2020Apr}, where networks with lower connectivity show higher plasticity and larger fracture strain. For same monomer concentration, finer networks formed at a higher temperature have smaller mesh size and the networks make more contacts with the shearing plates as compared to more bundled networks obtained at lower temperatures. This is illustrated in Fig. 5(c) and 5(d). This explains the increase of yield strain/stress with increasing temperature (Fig. 1(e)).
\newline 
\section{Conclusion}
In conclusion, we study the yielding and strain localization under steady shear in disordered, un-crosslinked networks of type-I collagen over a range of network architectures obtained by varying the polymerization temperatures and collagen concentrations. Using an unit cell based affine network model we attempt to generalize the yielding behaviour under different sample conditions in terms of microscopic network parameters. For rheology experiments, we use a cone-plate geometry that ensures homogeneous shear-stress field in the sample. This indicates that the observed strain localization does not originate due to any imposed stress heterogeneity in the system. We find that network weakening or yielding starts well below the fracture strain, signifying that yielding is a gradual process in disordered collagen networks. Remarkably, we find a strong correlation between the onset of network slippage and the yield strain for different imposed strain ramp rates, polymerization temperatures and collagen concentrations. Such observation points out that strain localization and boundary failure gives rise to yielding in collagen systems for a range of network architectures. Similar to earlier studies \cite{JANSEN20182665, Burla2020Apr}, we also observe that the networks with higher onset strain values for stiffening (observed for higher polymerization temperatures) also show higher values of the yield and breaking strain as indicated in Fig. 1(e). However, in contrary to the observation of network fracture at random locations \cite{Burla2020Apr}, we find that boundary slippage leads to detachment/rupture. Since, we image the entire gap between the shearing plates in-situ, our spatial resolution is only $\sim 25\, \mu$m which is much larger than the network mesh size \cite{Arevalo2010Oct}. This indicates that, although boundary failure results in the yielding and fracture of the bulk network, local plasticity and fracture over few mesh-sizes can not be ruled out. Furthermore, due to low spatial resolution, we can not confirm whether there is a slippage or, formation of a narrow band of high shear rate close the boundary. To resolve such issues, direct imaging of the network deformation in the flow-gradient plane using fluorescence microscopy or very fast z-scanning (along the gradient direction) of the network using a high speed laser scanning confocal microscope tracking the in-situ deformations, will be an interesting future direction to explore. These experiments can also directly probe number of contact formation with the shearing plates and can shed light on the local non-affinity giving rise to shear-thinning behaviour. 
Nevertheless, due to the roughness of the sand blasted plates, the velocity field very close to the boundary can be quite complex. Studying the failure dynamics by tuning the strength of adhesion of collagen networks with poly-L-lysine coated plates \cite{Nam2016May} will be very interesting. One important point to note is that freeze-fracture SEM provides more reliable values of filament/bundle diameters as it is not prone to the sample dehydration effect observed for dry SEM. However, we find that unlike dry SEM imaging, clear network images are obtained from freeze-fracture SEM only over small fields of view ($<< L_p$). Our observation of localized network rarefaction correlates well the yielding behavior. Nonetheless, more experimental and theoretical insights are needed to confirm such mechanism. Our study can provide strategies for delaying network failure and thus help in designing more resilient collagen based scaffolds for various engineering and biomedical applications. We hope that our work will motivate further studies on microscopic mechanism of failure and strain localization in collagen and other biopolymer networks.      

\section{Acknowledgments}
S.M. thanks SERB (under DST, Govt. of India) for a Ramanujan Fellowship. We acknowledge Ivo Peters for developing the Matlab codes used for PIV analysis. We thank Gautam Soni, Pramod Pullarkat, Reji Philip and Ranjini Bandyopadhyay for allowing us to use their lab/common instrument facilities. We also thank K M Yatheendran for help with the SEM imaging, Sachidananda Barik for synthesizing the polystyrene particles, Madhu Babu for the help with the FTIR measurements and RRI workshop facility for machining the humidity chamber for the sample.  

%%%END OF MAIN TEXT%%%


%\balance
\section{Materials and Methods}

\subsection{\textbf{Chemicals Used:}}
The collagen hydrogels are synthesised from acid-soluble rat tail collagen Type-I [Conc. 8.70 mg/ml in 0.02 N Acetic acid], (Corning, Bedford, MA). The collagen monomers are polymerized using the phosphate buffered saline (PBS) solution (1X, pH 7.34) prepared by uniform mixing of 2 g of NaCl, 50 mg of KCl, 0.36 g of Na$_2$HPO$_4$ and 60 mg of KH$_2$PO$_4$ in 250 mL of deionized water, with pH adjusted to 7.4 by adding 0.01 M HCl. The Polystyrene microbeads (PS) are used as the tracer particles are synthesized in the laboratory by a procedure described previously \cite{Dhar2019}.

\subsection{\textbf{Characterization of Collagen hydrogels:}}
\subsubsection{Rheology:}
For rheology experiments we use a 2018 made MCR-702 Twin-Drive stress-controlled rheometer (Anton Paar, Graz, Austria). A 25 mm diameter top cone geometry (cone angle: 2$^{\circ}$) with a 25 mm diameter parallel bottom plate (both made of stainless steel and both are sandblasted) are used. We mix the collagen monomers of a desired concentration with 1X PBS buffer and transfer the sample on the cold Peltier controlled rheometer bottom plate, immediately. Then the Peltier temperature is increased to the desired value. We polymerize collagen networks between the rheometer plates maintained at a fixed temperature. We also use a humidity chamber to prevent solvent evaporation during the rheology experiments. For in-situ imaging studies, we use a thin layer of 5 cSt silicone oil (Merck) around the sample to prevent solvent evaporation. 

\subsubsection{Freeze-fracture Electron Microscopy:} 
To probe the collagen network architecture, we carry out SEM imaging using an Ultra Plus Cryo-SEM (Zeiss, Germany) set-up. The collagen hydrogels of different concentrations (1 to 3 mg/ml) are prepared by adding the stock collagen to PBS solution in 1.5 ml centrifuge tubes, followed by a thorough mixing. Different samples of hydrogels were made by varying the polymerization temperature over the range 4$^{\circ}$ - 35$^{\circ}$C. Polymerized samples appear translucent under the ambient light. After polymerization, the samples were transferred to the sample holder for freeze-fracture SEM using a micro-pipette, as soon as possible. The sample holder is dipped in liquid nitrogen to freeze the samples instantly. Next, the frozen samples are mounted on the SEM sample stage using carbon tape and are sputter coated with platinum to a thickness of $\sim$ 5 nm using a Polaron SC7620 sputter coater (Watford, UK). After this, SEM images are recorded. The images are analyzed using ImageJ/Fiji/Matlab software.
 
\subsubsection{Fourier Transform Infrared Spectroscopy (FTIR):}
A Perkin-Elmer Spectrum 1000 FT-IR spectrometer is used to record the IR spectra of the dried samples of pristine PS, collagen hydrogel and collagen hydrogel with dispersed PS. The spectral positions are typically given in wavenumber ($cm^{-1}$) unit. The dried samples in very less quantity are mixed with KBr powder in a mortar and then ground well. The mixtures are then made into pellets, and loaded into the sample holder for measurements. Any new chemical interaction between collagen and PS should give rise to new troughs \cite{PayneFouriercollagen1988, VIDAL2011CollagenFourier, Belbachir2009Oct} for the composite samples as compared to the pristine samples of pure collagen and PS \cite{Fang2010Nov}. We find that almost all the troughs in the composite sample match with those for the individual pristine samples (Fig. S5). This indicates that the tracer PS do not form any chemical bonds with the collagen networks.

\subsection{Details of PIV analysis:}
To get the instantaneous velocity profiles across the gap between the shearing plates, we use particle imaging velocimetry (PIV) technique. PIV analysis is performed using custom written PIV codes developed using Matlab software. We first record a sequence of coloured images of the speckle pattern obtained from the in-situ imaging (Fig. S7)  at a fixed frame rate. Next we convert them into 8-bit grey-scale images. Then for a given image we divide the region of interest (ROI; 760 pix $\times$ 958 pix) into small square grids (each 40 pix $\times$ 40 pix). The spatial resolution is $\sim$ 0.6 $\mu$m/pix. We select the ROI such that we are at least 40 pixels away from the visible plate boundaries to avoid the roughness scale (obtained from optical imaging) of the sand blasted plates (RMS $\sim$ 25 $\mu$m). A typical PIV window is shown in Fig. 2a in the main text. Note that if the selected grid-size is too large, the spatial resolution of PIV analysis is reduced. If it is too small, there can be artefacts due to lack of averaging that introduces significant measurement errors. We find that for our experimental set-up, 40 pix $\times$ 40 pix is the minimum grid size required for reliable analysis. For larger grid sizes the measured velocity profiles remain similar but the resolution decreases, as mentioned earlier.

To obtain the local velocities in the flow-gradient plane (flow along x, gradient along z), as indicated in Fig. 2(a), we consider two consecutive images obtained at times $t$ and $t + \Delta t$. We match the intensity distribution inside a grid at time $t$ with that at time $t + \Delta t$ using cross-correlation technique to find the displacements inside the individual grids over the entire ROI. We obtain the local spatial displacements $\Delta x(x, z)$ and $\Delta z(x, z)$ over time $\Delta t$. From this the local velocities are obtained: $v_x (x, z) = \Delta x (x, z)/\Delta t$ and $v_z (x, z) = \Delta z (x, z)/\Delta t$. We find that for a given z position, the variation of $v_x (x, z)$ is negligible with respect to $x$. Thus, we can average over all $x$: $\left\langle v_x (x, z)\right\rangle _x = v_x (z)$. Each line in Fig. 2(b) corresponds to $v_x (z)$ vs $z$ obtained for different strain values. Also, we find that for all our experiments, $v_z (x, z) << v_x (x, z)$. 


%%%REFERENCES%%%
%\bibliography{ref} %You need to replace "rsc" on this line with the name of your .bib file
%\bibliographystyle{rsc} %the RSC's .bst file

\providecommand*{\mcitethebibliography}{\thebibliography}
\csname @ifundefined\endcsname{endmcitethebibliography}
{\let\endmcitethebibliography\endthebibliography}{}

\newpage
\textbf{\large{Supplementary Information: Strain localization and yielding dynamics in disordered collagen networks}}% Force line breaks with \\
%\thanks{Footnote to title of article.}
\vspace{3 cm}
%\author{Swarnadeep Bakshi}%
 %\affiliation{}%Lines break automatically or can be forced with \\

%\author{Vaisakh VM}
%\affiliation{} %This line break forced with \textbackslash\textbackslash\\

%\author{Ritwick Sarkar}
 %\homepage{http://www.Second.institution.edu/~Charlie.Author.}

%\author{Sayantan Majumdar}
% \homepage{http://www.Second.institution.edu/~Charlie.Author.}
%\thanks{smajumdar@rri.res.in}
%\affiliation{$^a$Soft Condensed Matter Group, Raman Research Institute, Bengaluru 560080, India\\ $^b$Department of Physics, HKUST, Clear Water Bay, Hong Kong}


%\begin{keyword}
    %collagen, fracture, network, rheology
%\end{keyword}
%\keywords{Collagen; yielding; fracture; rheology} %Use showkeys class option if keyword
%\maketitle

%\begin{quotation}
%The ``lead paragraph'' is encapsulated with the \LaTeX\ 
%\verb+quotation+ environment and is formatted as a single paragraph before the first section heading. 
%(The \verb+quotation+ environment reverts to its usual meaning after the first sectioning command.) 
%Note that numbered references are allowed in the lead paragraph.
%The lead paragraph will only be found in an article being prepared for the journal \textit{Chaos}.
%\end{quotation}

{\Large{Movie Descriptions}}
\\\\
\textbf{Movie1:}\,\,\,\,In this movie, we show the in-situ deformation of the sample surface in flow-gradient plane correlating with stress vs strain response of collagen network (2 mg/ml) seeded with 1\% (v/v) PS. The applied strain ramp rate is 1\%/s. The images are captured using a digital camera (Lumenera) fitted with a 5X, long working distance objective (Mitutoyo) at a frame rate of 1 Hz with a resolution of 1200 X 1800 (pixel)$^2$. The images are then cropped and compressed such that the spatial resolution is $\sim$1.4 $\mu$m/pixel. The movie is sped up 14 times as compared to the real time. As mentioned in the main text, we see that network slippage with one of the shearing plates (here, top plate) appears well before the peak stress/breaking stress is reached. The onset of such slippage correlates well with the yield strain of the network (as indicated in the movie). Beyond the peak stress, the detachment (characterized by an elastic retraction of the whole network) takes place from the top plate. The drop in scattered intensity near the top plate close to the yield strain indicates network rarefaction (also see Fig. S8 and the main text) before detachment.
\newline
\newline
\textbf{Movie2:}\,\,\,\,In this movie, we show the in-situ deformation of collagen network (2 mg/ml) seeded with 1\% (v/v) PS for an applied strain ramp rate of 10\%/s. The movie is sped up 1.4 times as compared to the real time. Here also, we find similar correlation of the boundary dynamics with the yield and breaking strains of the network (data not shown). Interestingly, as opposed to Movie1, the network detachment happens from both the plates in this case. Also, we see a clear signature of network rarefaction (drop in scattered intensity) near the shearing boundaries close to yielding (also see Fig. S8). Beyond the network breakage/rupture from both the plates, a clear signature of network contraction away from the plates is seen.  
\newpage

{\Large{Table of parameters}}

\begin{center}
Parameters obtained from the 8-chain model fitting for three different collagen concentrations.
%\newline
\newline
$\phi=$1mg/ml%\newline

\begin{tabularx}{0.8\textwidth}{ 
  | >{\centering\arraybackslash}X 
  | >{\centering\arraybackslash}X 
  | >{\centering\arraybackslash}X  
  | >{\centering\arraybackslash}X |}
 \hline
 Temperature ($^{\circ}$C) & n$k_B$T(J$m^{-3}$) & $\frac{L_p}{2L_{c}}$ & $\frac{\xi}{L_{c}}$\\
 \hline
 20  & 80  & 0.421 & 0.760\\
\hline
 25  & 180  & 0.400 & 0.738\\
\hline
 30  & 200  & 0.235 & 0.488\\
\hline
 35  & 270  & 0.234 & 0.479\\
\hline
 37  & 380  & 0.229 & 0.470\\
\hline
\end{tabularx}
\end{center}

\begin{center}
$\phi=$2mg/ml

\begin{tabularx}{0.8\textwidth}{ 
  | >{\centering\arraybackslash}X 
  | >{\centering\arraybackslash}X 
  | >{\centering\arraybackslash}X  
  | >{\centering\arraybackslash}X |}
 \hline
 Temperature ($^{\circ}$C) & n$k_B$T(J$m^{-3}$) & $\frac{L_p}{2L_{c}}$ & $\frac{\xi}{L_{c}}$\\
 \hline
 20  & 100  & 0.390 & 0.750\\
\hline
 25  & 335  & 0.366 & 0.717\\
\hline
 28  & 350  & 0.306 & 0.646\\
\hline
 30  & 370  & 0.284 & 0.606\\
\hline
 35  & 470  & 0.252 & 0.554\\
\hline
\end{tabularx}
\end{center}%\newline

%\newline
\begin{center}

$\phi=$3mg/ml

\begin{tabularx}{0.8\textwidth}{ 
  | >{\centering\arraybackslash}X 
  | >{\centering\arraybackslash}X 
  | >{\centering\arraybackslash}X  
  | >{\centering\arraybackslash}X |}
 \hline
 Temperature ($^{\circ}$C) & n$k_B$T(J$m^{-3}$) & $\frac{L_p}{2L_{c}}$ & $\frac{\xi}{L_{c}}$\\
 \hline
 20  & 300  & 0.463 & 0.790\\
\hline
 25  & 350  & 0.308 & 0.658\\
\hline
 28  & 380  & 0.273 & 0.607\\
\hline
 30  & 420  & 0.255 & 0.579\\
\hline
 35  & 500  & 0.253 & 0.574\\
\hline
\end{tabularx}
\end{center}
\newpage

{\Large{Supplementary figures}}
\begin{figure} [h]
    \begin{center}
    %trim from left edge
    \includegraphics[height = 6 cm]{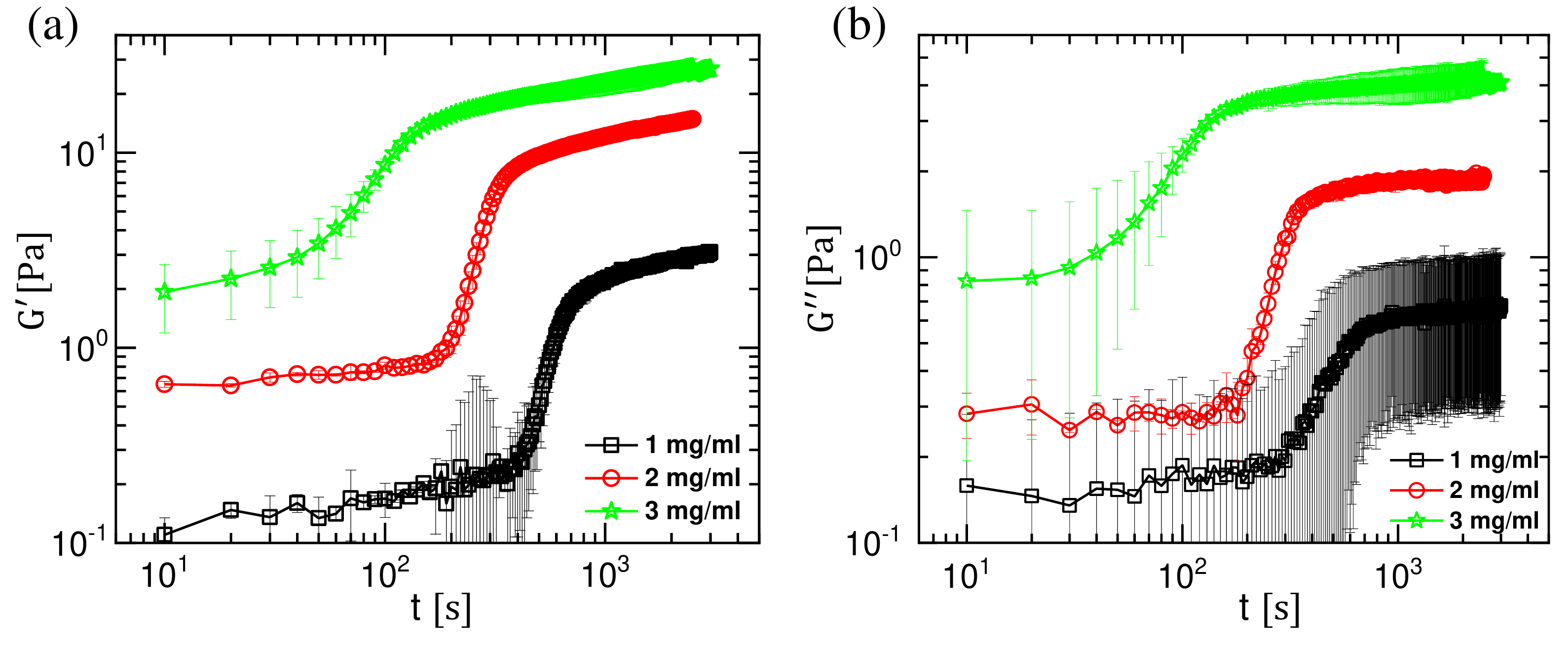}
		\renewcommand{\thefigure}{S1}
    %trim from right edge
    %\includegraphics[]{F1.PDF}
		\vspace*{-3mm}
    \caption{Variation of storage ($G'$, panel (a)) and  loss ($G''$, panel (b)) moduli as a function of time during the polymerization of collagen networks. The applied oscillatory strain amplitude is 2\% and frequency is 0.5 Hz. As indicated, different symbols represent different concentrations of collagen. The plateau reached after the jump in $G'$ or $G''$ values represents the polymerized state of the network. The error bars are the standard deviations of two independent measurements under the same condition. Here, polymerization temperature (T) is 25$^{\circ}$C in all cases.}
    \label{S1}
    \end{center}
\end{figure}
\begin{figure}[h]
    \begin{center}
    %trim from left edge
    \includegraphics[height = 6.2 cm]{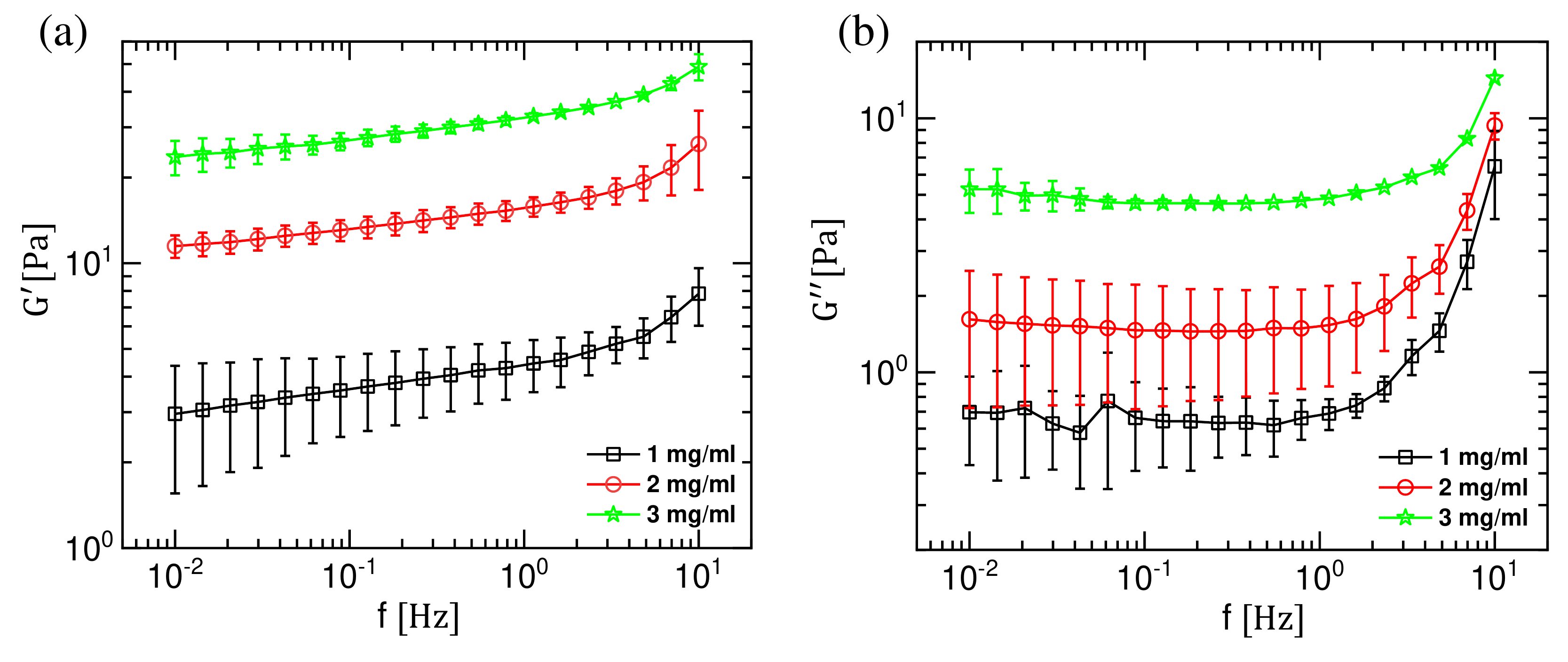}
		\renewcommand{\thefigure}{S2}
    %trim from right edge
    %\includegraphics[]{F1.PDF}
		\vspace*{-3mm}
    \caption{Frequency dependent storage ($G'$, panel (a)) and loss ($G''$, panel (b)) moduli for polymerized collagen networks. The applied oscillatory strain amplitude is 2\%. Different symbols represent different concentrations of collagen. T = 25$^{\circ}$C. The error bars are the standard deviations of two independent measurements under the same condition. We see that, $G' >> G''$ over the entire range of frequencies probed.}
    \label{S2}
    \end{center}
\end{figure}
\begin{figure}[h]
    \begin{center}
    %trim from left edge
    \includegraphics[height = 6.2 cm]{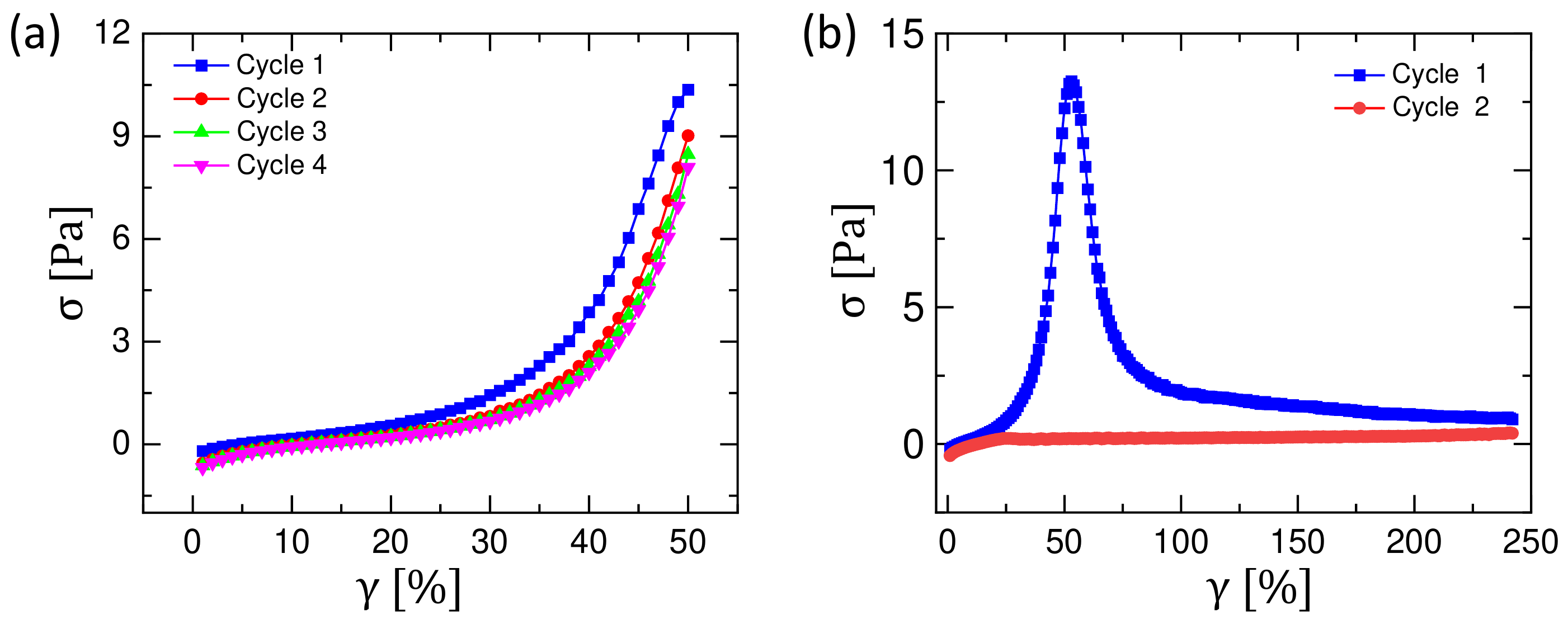}
		\renewcommand{\thefigure}{S3}
    %trim from right edge
    %\includegraphics[]{F1.PDF}
		\vspace*{-3mm}
    \caption{Stress ($\sigma$) vs strain ($\gamma$) response under repeated deformations for maximum strain below the yield strain (panel (a)) and well beyond yield/breaking strain (panel (b)). Below yield strain ($\gamma < \gamma_y$) the mechanical response is reversible, but above yielding due to breakage of contact with the shearing plates, the response becomes very weak and irreversible. Here, $\phi$ = 2 mg/ml and T = 25$^{\circ}$C.}
    \label{S3}
    \end{center}
\end{figure}
\begin{figure}[h]
    \begin{center}
    %trim from left edge
    \includegraphics[height = 6.5 cm]{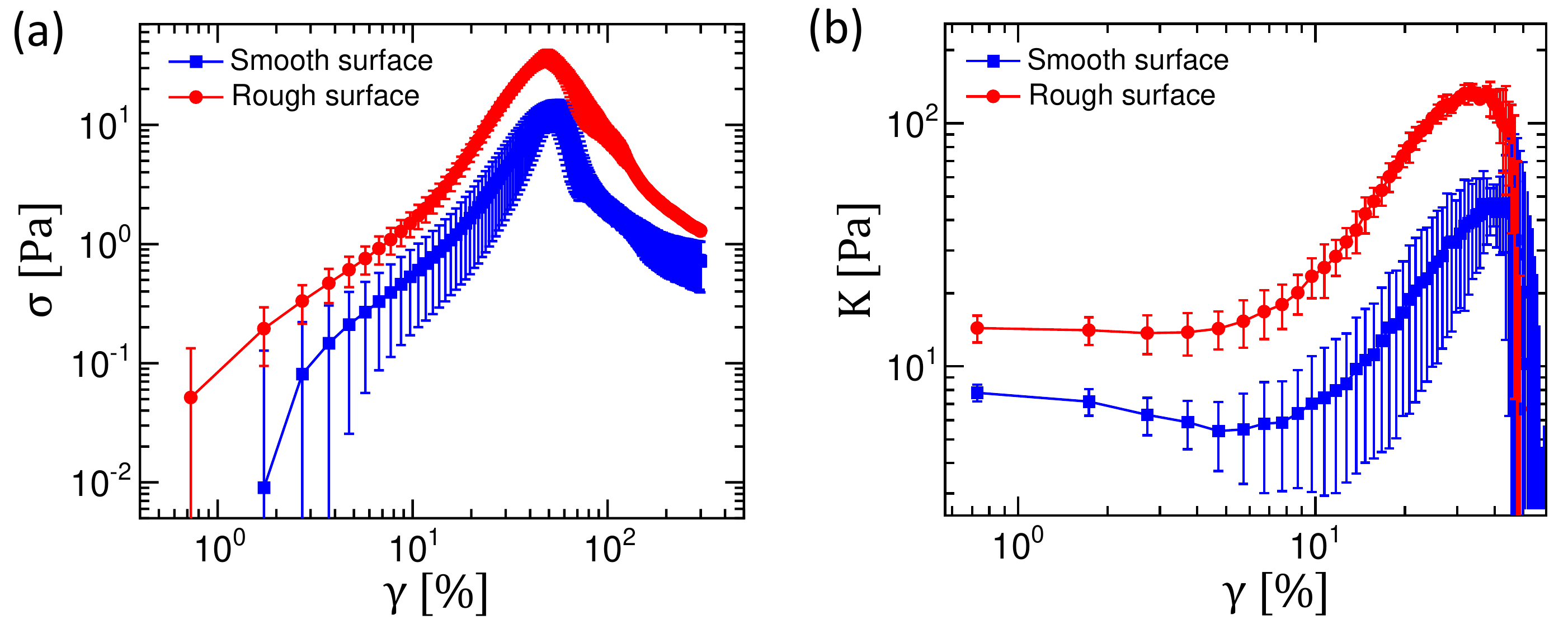}
		\renewcommand{\thefigure}{S4}
    %trim from right edge
    %\includegraphics[]{F1.PDF}
		\vspace*{-3mm}
    \caption{(a) Stress ($\sigma$) vs strain ($\gamma$) behaviour obtained from cone-plate geometries with smooth and rough (sand-blasted) surfaces as indicated. (b) Corresponding shear-moduli ($K$) vs $\gamma$. The standard deviations of four independent runs under the same condition are shown as error bars. Although, the rheolocical responses show similar trend, the absolute values of $K$ remain significantly lower for smooth geometry. Here, $\phi$ = 2 mg/ml and T = 25$^{\circ}$C.}
    \label{S4}
    \end{center}
\end{figure}
\begin{figure}[h]
    \begin{center}
    %trim from left edge
    \includegraphics[height = 7 cm]{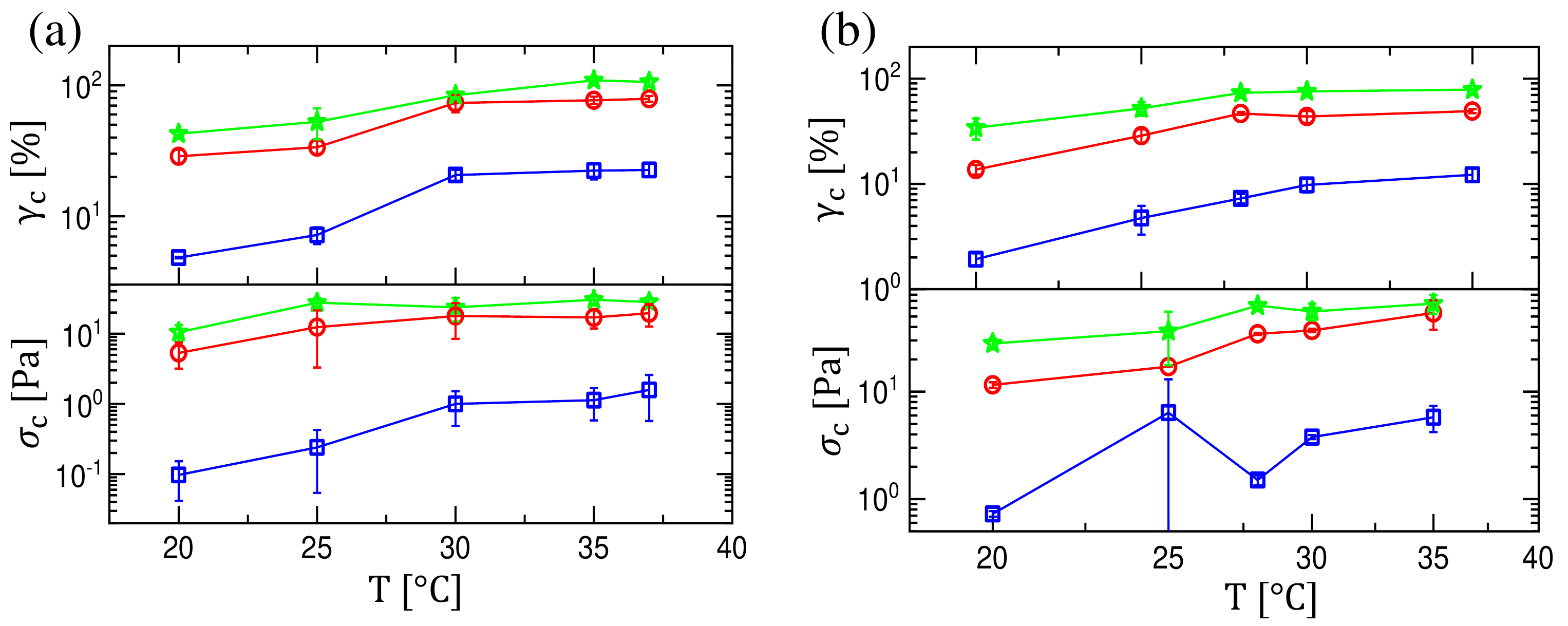}
		\renewcommand{\thefigure}{S5}
    %trim from right edge
    %\includegraphics[]{F1.PDF}
		\vspace*{-3mm}
    \caption{Variation of critical stress and strain values with polymerization temperature for collagen networks. Collagen concentration, $\phi$ = 1 mg/ml (panel (a)) and $\phi$ = 3 mg/ml (panel (b)). Here, squares represent onset strain/stress, circles represent yield strain/stress and stars represent breaking strain/stress. The error bars are the standard deviations of two independent measurements under the same condition.}
    \label{S5}
    \end{center}
\end{figure}
\begin{figure}[h]
    \begin{center}
    %trim from left edge
    \includegraphics[height = 7.5 cm]{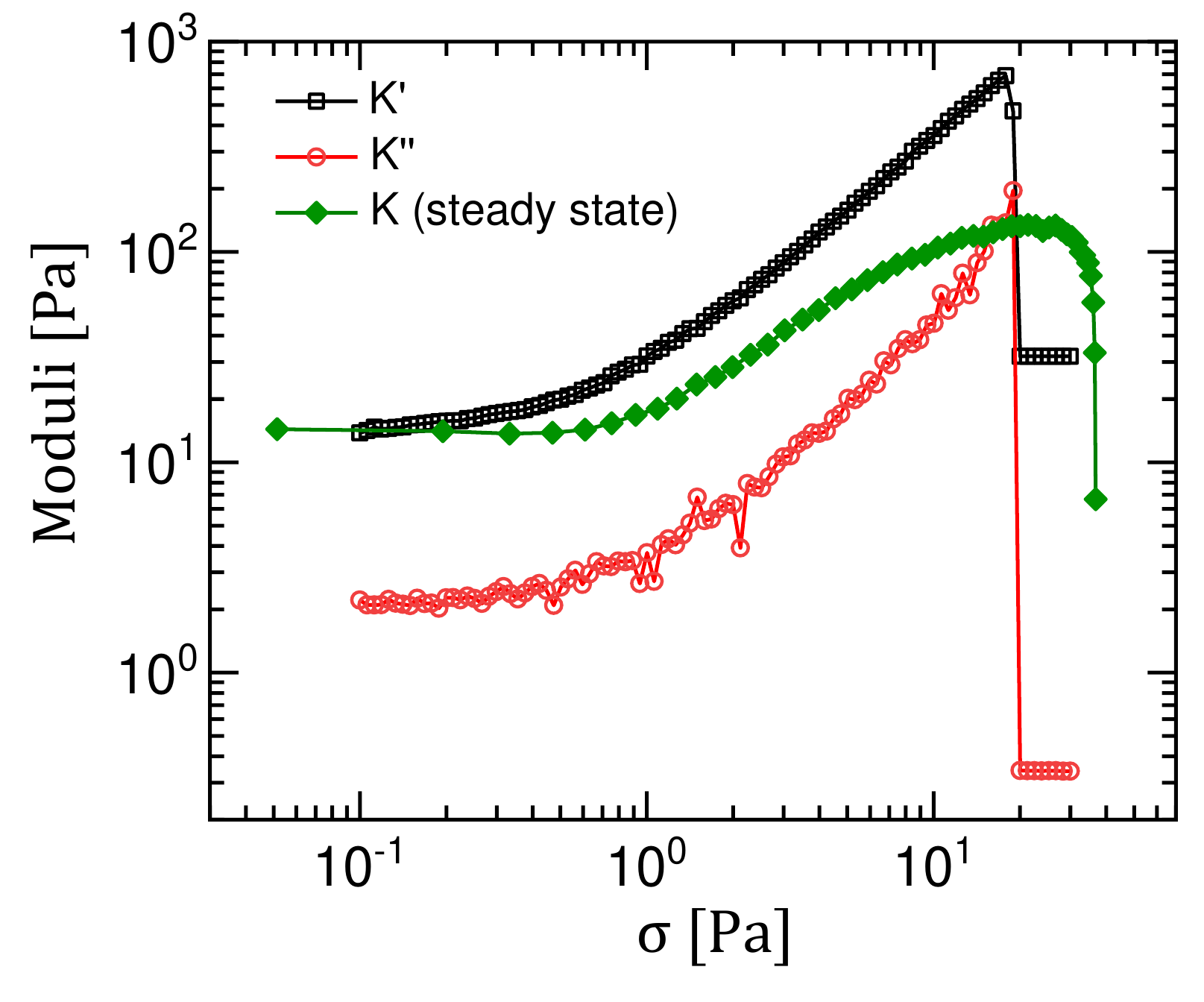}
		\renewcommand{\thefigure}{S6}
    %trim from right edge
    %\includegraphics[]{F1.PDF}
		\vspace*{-3mm}
    \caption{Storage ($K'$) and ($K''$) components of tangent shear modulus as a function of pre-stress ($\sigma$). $K'$ and $K''$ are obtained from superposing a small sinusoidal stress component (amplitude: 0.1 Pa and frequency: 1 Hz) on the D.C. pre-stress. Also, we compare these with the values of shear moduli ($K$) under steady state measurement under controlled strain condition (but plotted as a function of stress). We observe that there is a drop in $K'$, $K''$ and $K$ values after a strain stiffening response signifying yielding/network-rupture. Here, $\phi$ = 2 mg/ml and T = 25$^{\circ}$C.}
    \label{S6}
    \end{center}
\end{figure}
\begin{figure}[h]
    \begin{center}
    %trim from left edge
    \includegraphics[height = 9 cm]{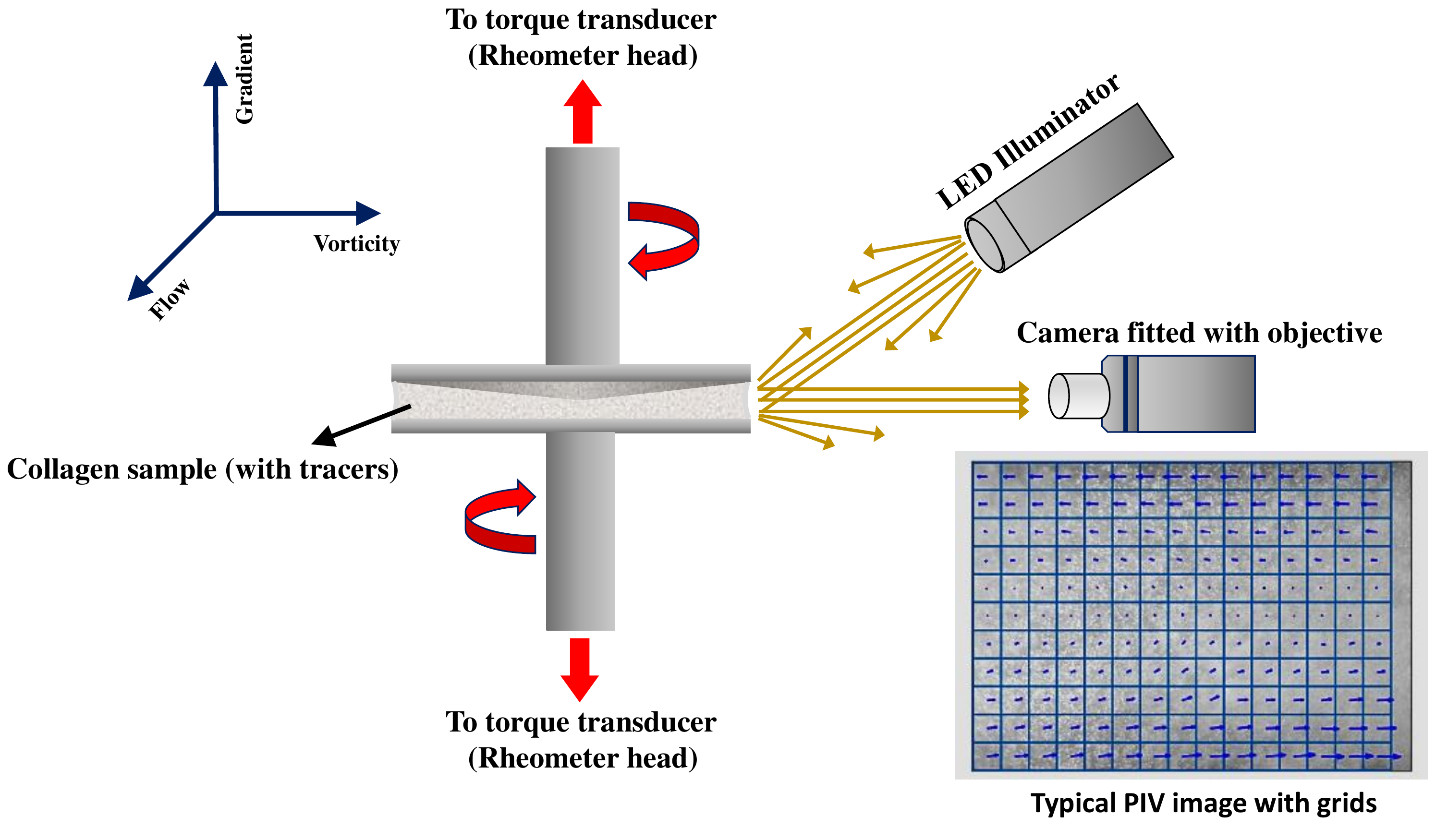}
		\renewcommand{\thefigure}{S7}
    %trim from right edge
    %\includegraphics[]{F1.PDF}
		\vspace*{-3mm}
    \caption{Schematic of the experimental set-up for rheology and in-situ optical imaging. A typical boundary image with superposed grids for PIV analysis is also shown.}
    \label{S7}
    \end{center}
\end{figure}
\begin{figure}[h]
    \begin{center}
    %trim from left edge
    \includegraphics[height = 5.5 cm]{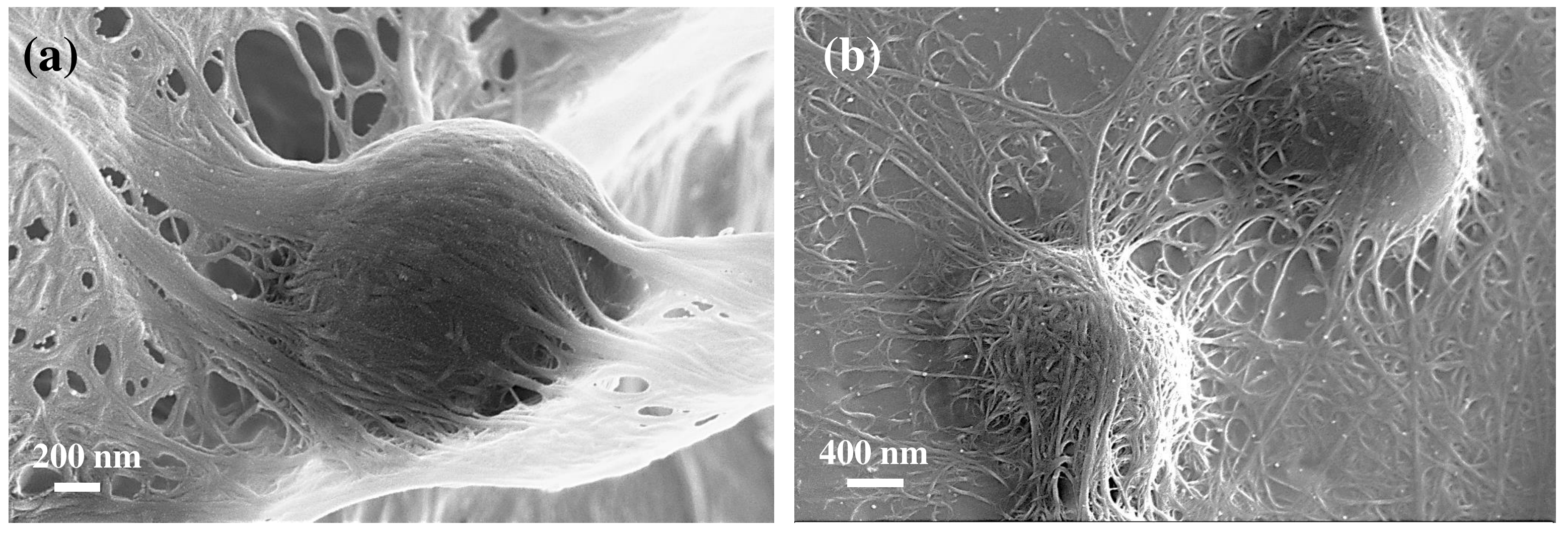}
		\renewcommand{\thefigure}{S8}
    %trim from right edge
    %\includegraphics[]{F1.PDF}
		%\vspace*{-3mm}
    \caption{Freeze fracture SEM image of collagen network ($\phi$ = 2 mg/ml) seeded with 1\% PS of mean diameter of 2.8 $\mu$m. The collagen fibrils seem to have some affinity to stick to the PS surface.}
    \label{S8}
    \end{center}
\end{figure}
\begin{figure}[h]
    \begin{center}
    %trim from left edge
    \includegraphics[height = 6 cm]{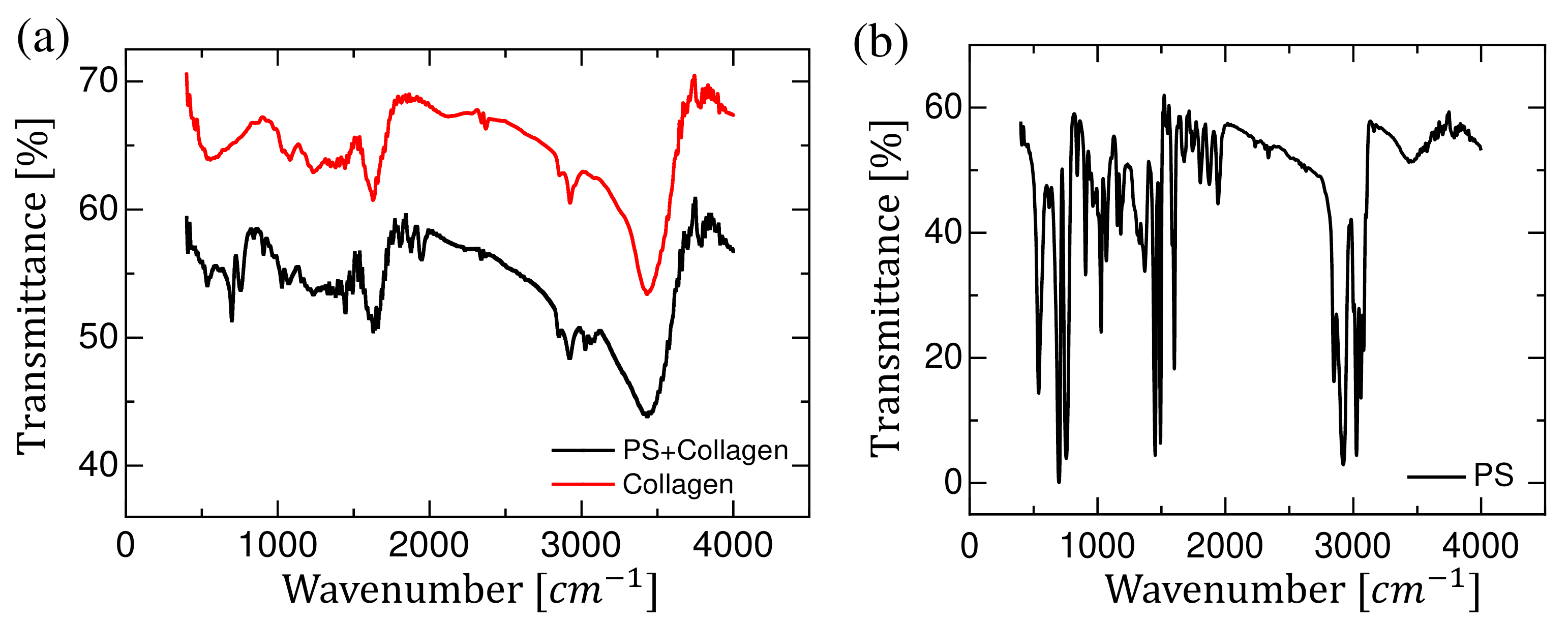}
		\renewcommand{\thefigure}{S9}
		\vspace*{-3mm}
    %trim from right edge
    %\includegraphics[]{F1.PDF}
    \caption{FTIR spectra of pure collagen (shown by the red line in (a)) and collagen mixed with 1\% PS (shown by the black line in (a)). FTIR spectrum for only PS particles are shown in (b). As indicated in (a), troughs shown in both the spectra (pure collagen and collagen + PS) are very similar. This indicates that there are no chemical bonding interactions between collagen and PS.}
    \label{S9}
    \end{center}
\end{figure}
\begin{figure}[h]
    \begin{center}
    %trim from left edge
    \includegraphics[height = 6.5 cm]{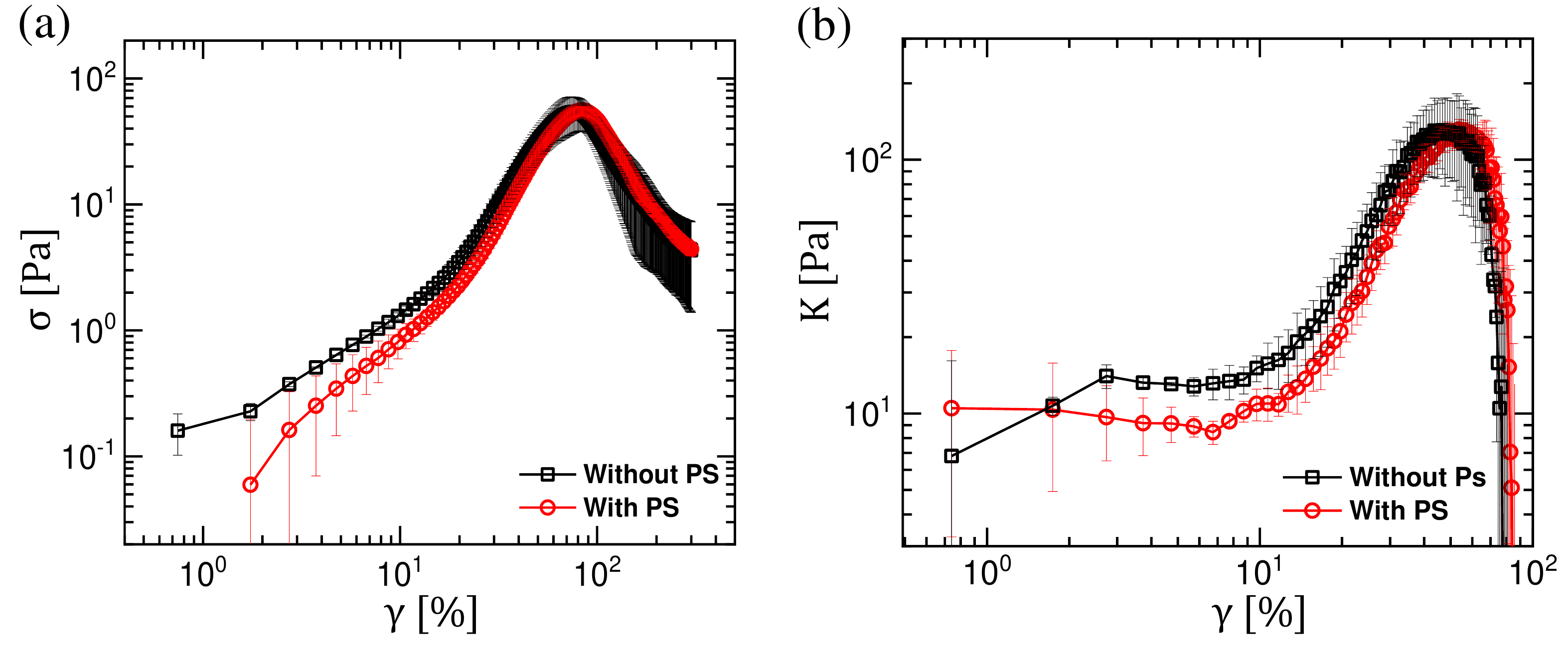}
		\renewcommand{\thefigure}{S10}
    %trim from right edge
    %\includegraphics[]{F1.PDF}
		\vspace*{-3mm}
    \caption{Comparison of network response with and without added PS (1\%). We see that stress ($\sigma$, panel (a)) and differential shear modulus ($K$, panel (b)) show very similar behaviour with applied strain ($\gamma$) for both the cases. The error bars are the standard deviations of two independent measurements under the same condition. Here, $\phi$ = 2 mg/ml and T = 30$^{\circ}$C.}
    \label{S10}
    \end{center}
\end{figure}
\begin{figure}[h]
    \begin{center}
    %trim from left edge
    \includegraphics[height = 7 cm]{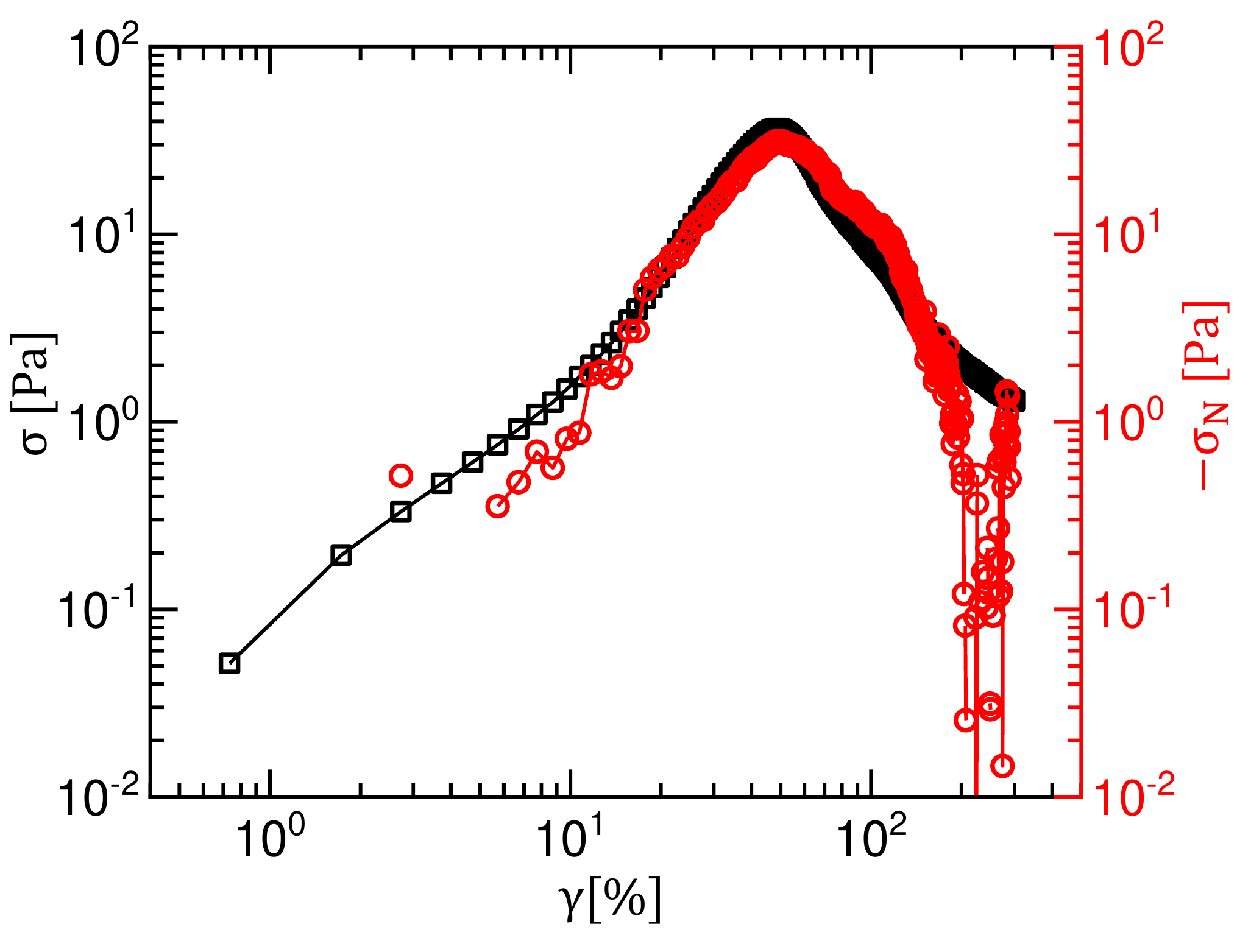}
		\renewcommand{\thefigure}{S11}
    %trim from right edge
    %\includegraphics[]{F1.PDF}
		\vspace*{-3mm}
    \caption{Typical variation of shear stress ($\sigma$, shown by the black squares) and negative normal stress ($\sigma_N$, shown by the red circles) as a function of applied strain $\gamma$ for collagen network with $\phi$= 2 mg/ml and T= 25$^{\circ}$C. We see that significant negative normal stress (comparable to the shear stress) develops in the network in the strain stiffening regime and reaches a maximum near the yielding.}
    \label{S11}
    \end{center}
\end{figure}
\begin{figure}[h]
    \begin{center}
    %trim from left edge
    \includegraphics[height = 10 cm]{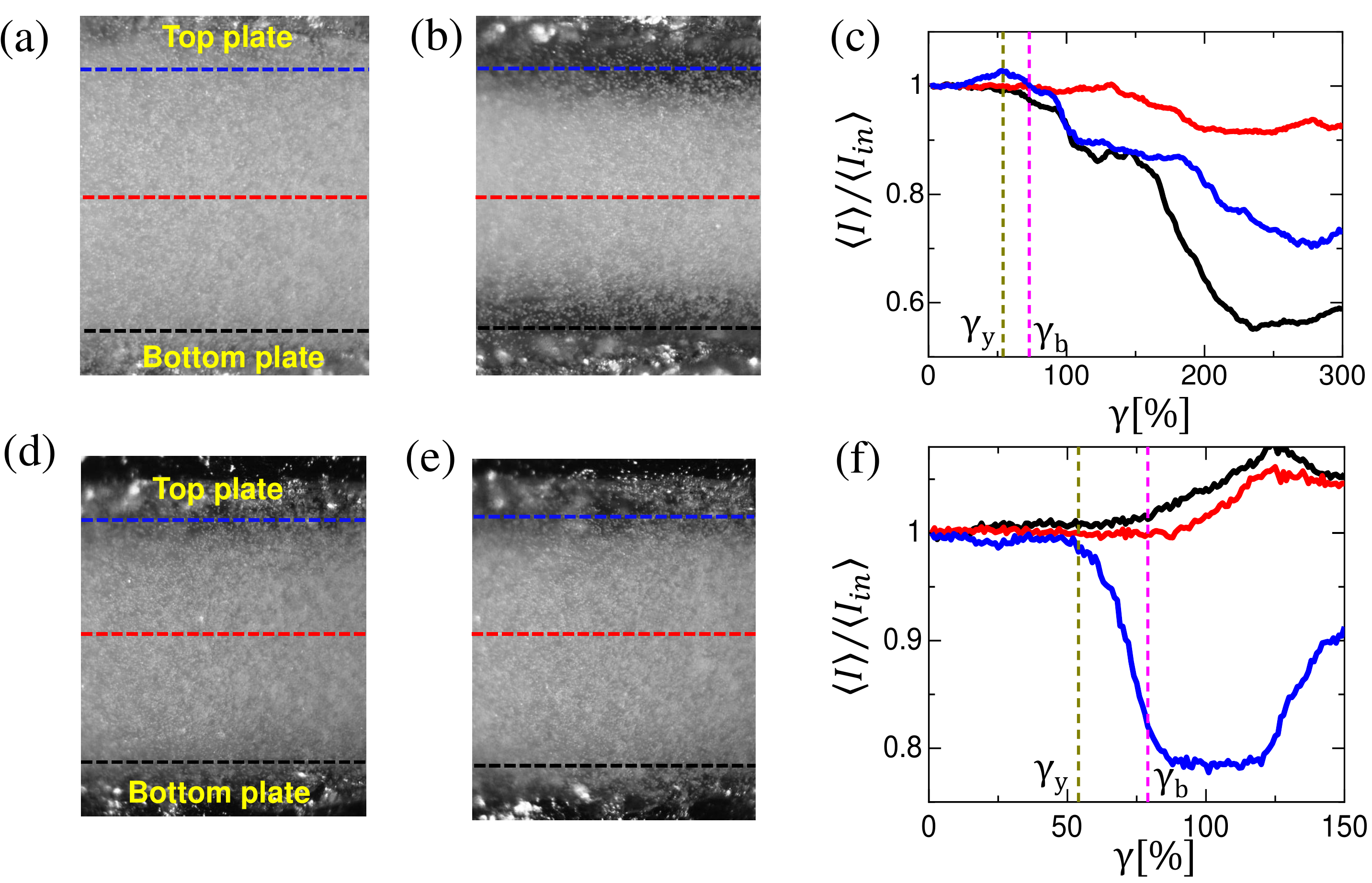}
		\renewcommand{\thefigure}{S12}
    %trim from right edge
    %\includegraphics[]{F1.PDF}
		\vspace*{-3mm}
    \caption{Panels (a), (b), (d) and (e) show typical boundary images of the sample with dashed horizontal lines parallel to the plates indicating the positions where the average intensity (normalized by the initial intensity) as a function of applied strain ($\gamma$) are calculated as shown in (c) and (f). Positions near the top and bottom plates indicated by blue and black lines, respectively. Red line indicates a position near the midway between the plates. (a), (b), (c) correspond to a strain ramp rate of 10\%/s and (d), (e), (f) correspond to 1\%/s. For both ramp rates, the intensity near both the plates (c) or, one of the plates (f) drops significantly beyond the yield ($\gamma_y$) and the breaking ($\gamma_b$) strains as indicated. Panels (a) and (d) correspond to the initial unstrained state ($\gamma$ = 0) of the sample, whereas, (b) and (e) represent that after the network rupture. The drop in scattered intensity near yielding indicates network rarefaction leading to detachment. Here, $\phi$ = 2 mg/ml and T = 30$^{\circ}$C.   
		}
    \label{S12}
    \end{center}
\end{figure}

\begin{figure}[h]
    \begin{center}
    %trim from left edge
    \includegraphics[height = 6 cm]{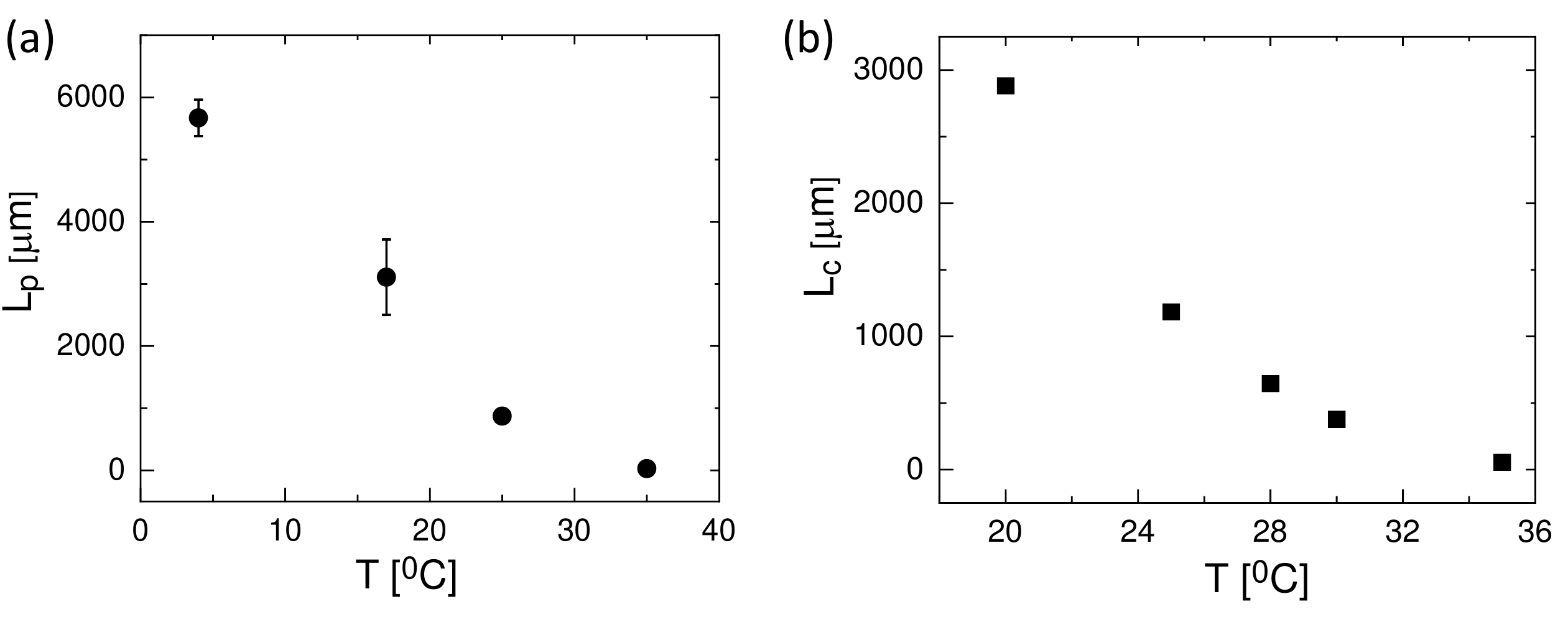}
		\renewcommand{\thefigure}{S13}
    %trim from right edge
    %\includegraphics[]{F1.PDF}
		\vspace*{-3mm}
    \caption{(a) Average persistence length ($L_p$) estimated from the mean bundle diameters obtained from freeze-fracture SEM data (Fig. 1c, main text) as a function of polymerization temperature. The error bars are the standard deviations obtained from the diameter distributions (Fig. 1c, main text). (b) $L_c$ values (described in the main text) obtained from the 8-chain model using the data for $L_p$ obtained from panel (a). We use the interpolated (cubic-spline) values of $L_p$ for estimating $L_c$ values at the temperatures shown in panel (b). Here, $\phi$ = 1 mg/ml.}   
		    \label{S13}
    \end{center}
\end{figure}

%\bibliography{PRM}
%\bibliographystyle{unsrt}

\end{document}